\documentclass[11pt,a4paper]{article}

\def\half{\frac{1}{2}}
\usepackage{axodraw}
\usepackage{jheppub}
\usepackage{graphics}

\author[a]{B.~C.~Allanach}
\affiliation[a]{Department of Applied Mathematics and Theoretical Physics, Centre for Mathematical Sciences, University of Cambridge, Wilberforce Road, Cambridge CB3 0WA, United Kingdom}
\emailAdd{B.C.Allanach@damtp.cam.ac.uk}
\author[b]{M A Parker}
\affiliation[b]{Cavendish Laboratory, University of Cambridge, JJ Thomson
  Avenue, Cambridge CB3 0HE, United Kingdom}
\emailAdd{parker@hep.phy.cam.ac.uk}

\title{Uncertainty in Electroweak Symmetry Breaking in 
  Models With High Scale Supersymmetry Breaking and its Impact on
  Interpretations of Searches For Supersymmetric  Particles}  

\abstract{Some regions of parameter space of the minimal supersymmetric 
  standard model (MSSM) with high scale supersymmetry breaking
  have extreme sensitivity of electroweak symmetry 
  breaking (EWSB) to the top quark mass through renormalisation group
  evolution effects. This leads to uncertainties in the
  predictions which need to be taken into account in the 
  interpretation of searches for supersymmetric particles in these regions. 
  As an example, we provide estimates of the current uncertainties on
  the position in parameter space of the region which does not break electroweak
  symmetry in the constrained MSSM (CMSSM). 
  The position of the boundary of EWSB can vary 
  by up to 2 TeV in $m_0$ due to the  
  uncertainties coming from the current measurement errors on 
  the top quark mass and from higher order corrections. In this dangerous
  region, for fixed CMSSM parameters the neutralino lightest supersymmetric
  particle mass has an associated large  
  uncertainty of order 100$\%$.  
  These uncertainties therefore have a profound effect on
  the interpretation of LHC supersymmetric particle searches in terms of the
  CMSSM\@. 
  We also show how to ameliorate poor
  convergence of the iterative numerical algorithm that calculates the MSSM
  spectrum near the boundary of EWSB\@. 
}

\begin{document}
\maketitle

\section{Introduction \label{sec:int}}
Searches for supersymmetric particles are one of the highest priorities for
the Large Hadron Collider, and the subject of many recent experimental and
theoretical publications. The data taken up until 2011 at a centre of mass
energy of 7 TeV has now been extensively analysed. New results have begun to
appear from 8 TeV data, and results from the complete 8 TeV data set are
expected to appear in 2013.  The most stringent limits on the masses of
supersymmetric squarks and gluinos come from  generic searches using the
missing energy signature, with a varying number of high transverse momentum
jets~\cite{Aad:2012rz, Chatrchyan:2012mfa}. The searches in
Ref.~\cite{Aad:2012rz} 
exclude gluino and squark masses below 860 GeV and 1320 GeV respectively in
simplified scenarios where only direct decays of these sparticles are
allowed and $R-$parity is assumed. In the constrained MSSM (CMSSM)
framework~\cite{Fayet:1,Fayet:2,Fayet:3,Fayet:4,Georgi:1}, where the sparticle
masses 
are fixed at a high gauge unification scale $M_X \sim 2 \times 10^{16}$ GeV,
the searches exclude universal gaugino masses
$M_{1/2}<640$~GeV at 
low values of universal scalar mass $m_0$, and $M_{1/2}<300$~GeV at high
$m_0$, approximately independently of the ratio of the two Higgs vacuum
expectation values $\tan \beta=v_2/v_1$ and the trilinear supersymmetry (SUSY)
breaking scalar coupling $A_0$.    
In the CMSSM
the limits approach the region where electroweak symmetry breaking (EWSB) does
not occur, and the search limits are generally not plotted beyond
the expected boundary of this region. The same can be seen in the most recent
conference interpretations of so far unpublished data from these
experiments. 

When the LHC runs at higher energy (13 or 14 TeV, starting in
2014), the searches will be sensitive to far higher values of $m_0$ and
$M_{1/2}$. It is 
therefore important to understand the behaviour of the theoretical predictions
as the boundary of EWSB is approached. The uncertainty on the location of the
boundary needs to be taken into account, as well as the uncertainties on the
predicted sparticle masses. 
%%%% Ben's bit added
These effects can significantly alter the region
of validity of the exclusion limits in interpretations of the data in terms of
high-scale SUSY breaking models (although
they of course do not alter the data themselves). 
In particular, if one is attempting to rule out the CMSSM (i.e.\ a bona fide
top-down analysis), the question of the
uncertainty of the position of the electroweak symmetry breaking boundary is a
physical one: in this case, we must fix the CMSSM parameters and ask if the
point is ruled out or not. This is also in principle the case for other
top down high-scale models of supersymmetry breaking where the various SUSY
breaking terms are set by a small number of parameters. 
If one is performing a `simplified model' or 
`phenomenological MSSM' approach,
one has enough free parameters to make sure that EWSB is broken correctly, 
and the problem should not arise. In practice, there is much interest from the
experiments in ruling out the CMSSM in particular because it is a familiar and
a priori reasonable model, and because it serves as a well-studied benchmark for
experimental progress in model space.

For a given point in MSSM parameter space, the computer program
{\tt SOFTSUSY}~\cite{Allanach:2001kg} is currently used by
ATLAS and CMS to calculate the sparticle spectrum and MSSM couplings, and to
determine whether or 
not the point performs EWSB successfully. We shall therefore also utilise 
{\tt SOFTSUSY} to quantify uncertainties due to the position of EWSB\@. Near the
boundary of successful EWSB, {\tt SOFTSUSY} reports non-convergence of the
numerical algorithm, which manifests as a small parameter region on the
experimental 
plots where the interpretation of an experimental search in terms of a
high-scale SUSY breaking model
cannot be trusted. 
Refs.~\cite{Feng:2011aa} and \cite{Matchev:2012vf} have also recently
noticed that at high values of third family scalar masses, sparticle spectrum
calculators (including {\tt SOFTSUSY}) give inaccurate
results: the sparticle mass calculation is particularly difficult to solve 
in this tricky region of parameter space. The region of parameter MSSM space
with large scalar masses 
has received considerable attention in the literature, and is called the focus
point, or hyperbolic branch~\cite{Chan:1997bi,Feng:1999mn,Feng:1999zg}.
The focus point has received renewed attention recently because if the SUSY
breaking scalar trilinear coupling involving Higgs and stops $A_t$ is large,
then a 
125 GeV lightest CP even Higgs boson (as implied by 2012 LHC
Higgs search data~\cite{a:2012gu,a:2012gk}) may
result~\cite{Feng:2012jf,Akula:2012kk,Nath:2012nh}.   

\subsection{EWSB in the MSSM}

We write the MSSM superpotential in terms of chiral superfields, suppressing
gauge indices and neglecting all Yukawa couplings save those of the third
family, 
\begin{equation}
W=\mu H_2 H_1 + h_t Q_3 H_2 u_3 + h_b Q_3 H_2 d_3 + h_\tau L_3 H_1 e_3,
\end{equation}
where the chiral superfields of the MSSM have the 
following $G_{SM}=SU(3)_c\times SU(2)_L\times U(1)_Y$ quantum numbers
\begin{eqnarray}
L_i:&(1,2,-\half),\quad {e}_i:&(1,1,1),\qquad\, Q_i:\,(3,2,\frac{1}{6}),\quad
{u}_i:\,({\bar 3},1,-\frac{2}{3}),\nonumber\\ {d}_i:&({\bar
  3},1,\frac{1}{3}),\quad 
H_1:&(1,2,-\half),\quad  H_2:\,(1,2,\half),
\label{fields}
\end{eqnarray}
and $i \in \{ 1,2,3\}$ is a family index. We shall write superpartners of SM
fields with a tilde. 

Minimising the MSSM Higgs potential with respect to the electrically neutral
components of the Higgs vacuum expectation values, one obtains the well-known
tree-level result for the Higgs mass parameter $\mu$ in the modified
dimensional reduction 
scheme $(\overline{DR})$
\begin{equation}
\mu^2 = 
 \frac{\tan 2\beta}{2} \left[m_{{H}_2}^2\tan \beta
- m_{{H}_1}^2  \cot \beta \right] - \frac{M_Z^2}{2}. \label{higgsMin}
\end{equation}
In order to reduce\footnote{This prescription at least ensures that the
  dominant terms do not involve large logarithms.} missing higher order
corrections, 
all quantities in Eq.~\ref{higgsMin} are understood to be evaluated at a
$\overline{DR}$ renormalisation scale $Q=M_{SUSY}$, where $M_{SUSY}$ is the
geometric mean of the two stop masses.  
$\tan \beta=\langle H_2^0 \rangle / \langle H_1^0 \rangle$ is the ratio of the
two MSSM Higgs vacuum expectation values
 and $m_{{H}_{1,2}}$ are the soft SUSY breaking $\overline{DR}$ mass terms of
 the Higgs doublets.
%If we rearrange Eq.~\ref{higgsMin} into a prediction for $M_Z$,
%then if $\mu \gg M_Z$ or $m_{H_2} \gg M_Z$, there must be a {\em a priori}
%unexpected large cancellation 
%between the term involving $\mu$ and the one involving $m_{H_2}$ in order to 
%provide a small enough value for $M_Z$. Such a cancellation constitutes a
%fine-tuning problem
If $m_{H_1}^2$ and $m_{H_2}^2$ and $\tan 2 \beta$ are such that $\mu^2>0$
results from Eq.~\ref{higgsMin}, the
model point may break electroweak symmetry successfully. On the other hand, if
$\mu^2 \le 0$, electroweak symmetry is {\em not}\/ broken successfully and the
model point is ruled out. We shall here study current uncertainties
in the position in 
MSSM parameter space of the $\mu=0$ contour that separates the region of
successful EWSB from the region of no EWSB\@. We shall also study the extent to
which these matter in terms of searches for supersymmetric particles.

The $m_{H_i}^2$ terms on the right hand side of Eq.~\ref{higgsMin} are
connected to the 
squark and gluino masses in models of SUSY breaking through the
boundary conditions on SUSY breaking terms, imposed at a (typically high)
scale $M_X$ 
and through the renormalisation group equations (RGEs), which at one-loop order
are~\cite{Martin:1993zk} 
 \begin{eqnarray}
16 \pi^2\frac{\partial m_{H_2}^2}{d t} &=& 
6 \left[ (m_{H_2}^2 + m_{{\tilde Q}_3}^2 +
  m_{{\tilde u}_3}^2 + A_t^2)  h_t^2  \right] - 6 g_2^2 M_2^2 - \frac{6}{5}
g_1^2 M_1^2 + \frac{3}{5}
g_1^2 \left(m_{H_2}^2-m_{H_1}^2 + \right. \nonumber \\
&&\left. \mbox{Tr}[m_{\tilde Q}^2 - m_{\tilde L}^2 - 2 
m_{\tilde u}^2 + m_{\tilde d}^2 + m_{\tilde e}^2] \right), \label{rgesA} \\
16 \pi^2\frac{\partial m_{H_1}^2}{d t} &=& 
6 \left[ (m_{H_1}^2 + m_{{\tilde Q}_3}^2 +
  m_{{\tilde d}_3}^2 + A_b^2)  h_b^2  \right] - 6 g_2^2 M_2^2 - \frac{6}{5}
g_1^2 M_1^2 -
\frac{3}{5}
g_1^2 \left(m_{H_2}^2-m_{H_1}^2 + \right. \nonumber \\
&&\left. \mbox{Tr}[m_{\tilde Q}^2 - m_{\tilde L}^2 - 2 
m_{\tilde u}^2 + m_{\tilde d}^2 + m_{\tilde e}^2] \right), \label{rges}
\end{eqnarray}
where $t=\ln Q$. 
When $(m_{H_2}^2 + m_{{\tilde Q}_3}^2 +
  m_{{\tilde u}_3}^2 + A_t^2)$ is large,
 the renormalisation of $m_{H_2}$ (and therefore its value at $M_{SUSY}$), is
 extremely sensitive to 
$h_t$, the top Yukawa coupling. $h_t$ is usually the largest dimensionless
coupling 
of the MSSM and is determined at the weak scale by the experimental input for
the top quark mass 
\begin{equation}
h_t(M_Z)=\frac{\sqrt{2}\ m_t(M_Z)}{{v(M_Z)} \sin \beta(M_Z)}, \label{ht}
\end{equation}
where $v(M_Z) \approx 246$ GeV is the Higgs vacuum expectation
value (VEV) that would be extracted for a single Higgs boson (in terms of the
two 
MSSM Higgs VEVs $v_{1,2}$, $v=\sqrt{v_1^2+v_2^2}$).
The running top mass $m_t(M_Z)$ is determined from the pole top mass
$m_t^{pole}$ by
\begin{equation}
 m_t(M_Z) = {m_t^{pole}} - \Delta m_t(M_Z), \label{topCorr}
\end{equation}
where $\Delta m_t(M_Z)$ are loop corrections to the top quark mass
evaluated at the
$\overline{DR}$ renormalisation scale $M_Z$.   $h_t$ is
around 1 for typical MSSM spectra (for instance, at parameter point
CMSSM10.1.1~\cite{AbdusSalam:2011fc}, 
$h_t(M_Z)=0.90$).  
Eqs.~\ref{rgesA},\ref{rges} are used, along with the other RGEs of the MSSM,
to evolve $m_{H_{1,2}}^2$ along with the other 
soft SUSY breaking parameters to $M_{SUSY}$, where the
Higgs potential is minimised and Eq.~\ref{higgsMin} is applied.
The region of successful EWSB is given by $\mu(M_{SUSY})^2 > 0$ in
Eq.~\ref{higgsMin}, i.e.\  (since $\tan \beta>1$ for models in which the
lightest CP even Higgs is heavy enough to satisfy experimental constraints,
$ \tan 2\beta<0$),
\begin{equation}
 \left[m_{{H}_2}^2 (M_{SUSY}) \tan \beta
- m_{{H}_1}^2(M_{SUSY})  \cot \beta \right] < \frac{M_Z^2}{\tan 2\beta}.
\label{ineq}
\end{equation}
Eq.~\ref{ineq} splits the MSSM parameter space into two regions:
one region has successful EWSB whereas the other does not. 

In models where we consider the primordial SUSY breaking terms to be set at
some higher scale $M_X > M_{SUSY}$, 
we shall now show that whether or not
this condition is satisfied depends upon the other soft SUSY breaking
parameters as well as $m_{H_2}^2(M_X)$, since $m_{H_i}^2(M_{SUSY})$ is
determined by solving Eqs.~\ref{rgesA},\ref{rges}.
To leading order in $A=\log (M_X/M_{SUSY})/(16 \pi^2)$, we solve
them to obtain 
\begin{eqnarray}
m_{H_2}^2(M_{SUSY}) &=& m_{H_2}^2(M_X) - A\left(
6 \left[ m_{H_2}^2(M_X) + m_{{\tilde Q}_3}^2(M_X) +m_{{\tilde u}_3}^2(M_X) +
\right. \right. \nonumber \\
&& \left. \left.   A_t^2(M_X)  \right] h_t^2(M_X) 
 \right). \label{approx1}\\
m_{H_1}^2(M_{SUSY}) &=& m_{H_1}^2(M_X) - A\left(
6 \left[ (m_{H_1}^2(M_X) + m_{{\tilde Q}_3}^2(M_X) +m_{{\tilde u}_3}^2(M_X) +
\right. \right. \nonumber \\
&& \left. \left. \left.  A_b^2(M_X) \right]  h_b^2(M_X)  \right]
 \right), \label{approx2}
\end{eqnarray}
where we have neglected terms proportional to electroweak gauge couplings
$g_{1,2}^2$ in favour of those proportional to $h_t^2$ or $h_b^2$.
This is a good approximation when $6 A h_{t,b}^2(M_X) \ll 1$, 
which is {\em not}\/ the case in the CMSSM\@. However, it is qualitatively 
instructive: 
substituting Eqs.~\ref{approx1},\ref{approx2} into Eq.~\ref{ineq}, we
obtain an inequality on the soft SUSY breaking parameters evaluated at $M_X$:
\begin{eqnarray}
m_{H_2}^2 &<& \frac{1}{1-6Ah_t^2} \left\{ 6 A h_t^2 \left( m_{{\tilde
        Q}_3}^2 + m_{{\tilde u}_3}^2 + A_t^2\right) +
\frac{M_Z^2}{\tan \beta \tan 2    \beta} + \right. \nonumber \\ && \left.
\frac{1}{\tan^2 \beta} \left[ m_{H_1}^2 -6 A h_b^2 (m_{{\tilde
        Q}_3}^2 + m_{{\tilde d}_3}^2 + A_b^2) \right]
\right\},
\end{eqnarray}
where $h_{t,b}$, $m_{H_2}^2$, $m_{{\tilde Q}_3}^2$, $m_{{\tilde u}_3}^2$,
$m_{{\tilde d}_3}^2$, $A_t$ and $A_b$ are all evaluated at $M_X$.
For large enough values of $m_{H_2}^2(M_X)$, we see that the inequality is not
satisfied and EWSB is {\em not}\/ broken successfully. Qualitatively, this
behaviour could also be 
analytically derived in the CMSSM by examining the complicated re-summed
logarithm solutions for
the soft masses in Ref.~\cite{Abel:1997va} in the limit that $h_b$ is also
neglected in the RGEs 
(although we do not include them here because they are 
not especially illuminating).

Much emphasis has been placed in the past upon fine-tuning associated with the
region close to the EWSB bound~\cite{Feng:1999mn,Feng:1999zg} where
$m_{H_2}^2(M_{SUSY})$ is somewhat insensitive to the value of $m_{H_2}^2(M_X)$
assumed. In Ref.~\cite{Allanach:2000ii}, a study of the uncertainties of the
position of the EWSB boundary in CMSSM parameter space was included,
explicitly showing its extreme sensitivity to the top quark
mass. Since then, however, the experimental precision with which the top mass
has been measured has been much improved, reducing the concomitant uncertainty. 
The time is ripe for a reanalysis of the uncertainties on EWSB, so that the
results may be used for future LHC searches. 
In section~\ref{sec:unc},
we shall find that the uncertainties are still large {\em despite}\/ the
improved precision upon the top mass measurement, and furthermore we shall 
demonstrate that they impact on the prospects of discovery. In particular, such
uncertainties should be included in the theoretical error when interpreting
experimental results in terms of high-scale SUSY breaking models. 
As a by-product of
our study, in section~\ref{sec:con}, we shall improve the numerical evaluation of the supersymmetric
spectrum in this tricky region of parameter space, as determined by {\tt
  SOFTSUSY3.3.4}. In section~\ref{sec:sum}, we summarise our results, as well
as provide some simple recommendations for interpretations of SUSY searches
in terms of high-scale SUSY breaking models that, if followed, will take the
uncertainties inherent in EWSB into account.  

\section{Uncertainties in Electroweak Symmetry Breaking \label{sec:unc}}

We now quantify the uncertainty inherent in the position of the boundary of
successful EWSB, and its effect on interpretations of searches for
supersymmetric 
particles in terms of high-scale SUSY breaking models. This will allow us to
suggest some practical steps for  
interpreters of data that allow such theoretical uncertainties to be
taken into account. We expect many models of SUSY
breaking in the MSSM to have regions of parameter space where EWSB is not
successfully broken. There are however some very constrained MSSM models of SUSY
breaking where EWSB 
is always successful (for example, the large volume string compactified
models~\cite{Balasubramanian:2005zx}, where the ratio of the universal scalar
mass to the 
universal gaugino mass is constrained to be $1/\sqrt{3}$), or phenomenological
models where one fixes $m_{H_2}^2$ independently of the other soft SUSY
breaking parameters, discarding the no EWSB region. However, for most models, 
we expect problematic regions of unsuccessful EWSB to exist, as in the CMSSM\@. 
Bad EWSB regions have been found
in minimal anomaly mediation, gaugino mediation and general gauge
mediation models to name just a few (see
Refs.~\cite{Barr:2002ex,Abel:2010vba,Yanagida:2013ah}, respectively). 
Here, we shall
focus for definiteness on the 
familiar example of the CMSSM, bearing in mind that very similar effects will
be present in most models of SUSY breaking. 
We expect that
our qualitative conclusions will hold in other models which possess
such 
regions in SUSY breaking parameters which are set at a high scale. At the very
least, our recommendations of how to quantify such 
uncertainties should be followed in any model which has a region of parameter
space which has no EWSB\@.

In the CMSSM, the focus point region is close to the region of unsuccessful EWSB
and shows extreme sensitivity to the top quark
mass through RGE effects. In
this region of CMSSM parameter space,  
$(m_{H_2}^2 + m_{{\tilde Qa}_3}^2 +   m_{{\tilde u}_3}^2 + A_t^2)$ is large in
Eq.~\ref{rgesA}, and $m_{H_2}(M_{SUSY})$ in Eq.~\ref{higgsMin} is very
sensitive to the value of $h_t$. As explained above, 
$h_t(M_Z)$ is in turn fixed by Eq.~\ref{ht}, via the 
running top mass $m_t(M_Z)$. This is affected by the pole mass input, which
has significant experimental errors (we take $m_t=173.5 \pm 1$
GeV~\cite{Beringer:1900zz}) as 
well as higher order corrections which are not included in the approximation.
However, if one fixes
SUSY breaking parameters at the electroweak scale, even if there are
constraints among the various parameters and a no-EWSB region exists, the
dominant uncertainties coming from the top mass (which enter via the RGE) are
not present. We note that the uncertainties originating from $m_t$ are
distinct from those that originate from two-loop (or higher) corrections to
the pole masses of supersymmetric particles~\cite{Allanach:2003jw} or higgs
bosons~\cite{Allanach:2004rh}, 
which are small (typically a few percent).

We include all
one-loop MSSM contributions to the top mass, as well as the two-loop QCD
corrections and two-loop RGE corrections.  
However, two loop SUSY QCD corrections to the top mass from squarks and
gluinos are {\em not}\/  taken
into account via Eq.~\ref{topCorr} in {\tt SOFTSUSY3.3.4}, nor are 3-loop RGE
corrections to $h_t$. 
The full MSSM two-loop ${\mathcal O}(\alpha_s^2)$ threshold corrections to
$m_t$ and three-loop RGE effects were calculated numerically in the
CMSSM~\cite{Bednyakov:2005kt,Bednyakov:2010ni} to be 
at the $2-3\%$ level for $\tan \beta=50$ and $A_0=0$. There, it was noticed
that in the focus point region, the position of the no EWSB forbidden region
moves by around 300 GeV in $m_0$.
In order to estimate the size of the ${\mathcal O}(\alpha_s^2)$ threshold
corrections for our sets of parameters, we   
use Eq.~(62) of Ref.~\cite{Bednyakov:2002sf}, which gives the two-loop SUSY QCD
contribution in the limit that the squark and gluino masses are all of equal
mass $M$: 
\begin{eqnarray}
\left(\frac{\Delta m_t(Q)}{m_t(Q)}\right)_{2LSQCD}&=&
\frac{4}{3} \left( \frac{\alpha_s}{4\pi} \right)^2
    \left\{ 
      \frac{47}{3} 
     + 20\ln\left(\frac{M^2}{Q^2}\right) 
     + 6\ln\left(\frac{M^2}{Q^2}\right) \ln\left(\frac{M^2}{m_t^2}\right)
      \right. \nonumber\\ 
&& +
 \frac{4}{3} \left[ \frac{23}{24} 
           - \frac{13}{6}\ln\left(\frac{M^2}{Q^2}\right) 
           + \frac{1}{2}\ln^2\left(\frac{M^2}{Q^2}\right)
       - 3\ln\left(\frac{M^2}{Q^2}\right) 
                     \ln\left(\frac{m_t^2}{Q^2}\right)
      \right] \nonumber\\ 
&& +
   3 \left[ 
      \frac{175}{72} + \frac{41}{6}\ln\left(\frac{M^2}{Q^2}\right) 
     - \frac{1}{2}\ln^2\left(\frac{M^2}{Q^2}\right) 
     - 2\ln\left(\frac{M^2}{Q^2}\right)\ln\left(\frac{m_t^2}{Q^2}\right)
       \right] \nonumber\\
&& + 
%AVB
  \frac{%-X_t+\mu\tan\beta
  X_t}{M} \left[
    - 4 - 8\ln\left( \frac{M^2}{Q^2}\right) \right] %\nonumber\\
%&& 
+ \frac{4}{3} \, \frac{%-X_t+\mu\tan\beta
  X_t}{M} \left[
    \frac73 - \frac{11}{3}\ln\left( \frac{M^2}{Q^2}\right) 
       + 3\ln\left( \frac{m^2_q}{Q^2}\right)  \right]\nonumber\\
&& +\left.
  3 \, \frac{%-X_t+\mu\tan\beta
  X_t
}{M} \left[
    - \frac83 
     + 4\ln\left( \frac{M^2}{Q^2}\right)  \right]
 \right\}, \label{limit}
\end{eqnarray}
where $X_t=A_t-\mu \cot \beta$.
The squark masses are typically much larger than the gluino masses
in the focus point 
region, and so this may 
be a bad approximation to the 
actual correction.
However, we may estimate the  size of
the theoretical error roughly by estimating the size of Eq.~\ref{limit} 
across parameter space
(i.e.\ assuming that 
the squark and gluino masses are all equal to the average of the dimensionally
reduced values of both stop masses and the gluino mass). 

\begin{figure}
\unitlength=1in
\begin{center}
\begin{picture}(3,2.5)
  \put(-0.7,3){\includegraphics[angle=270,width=0.8\textwidth]{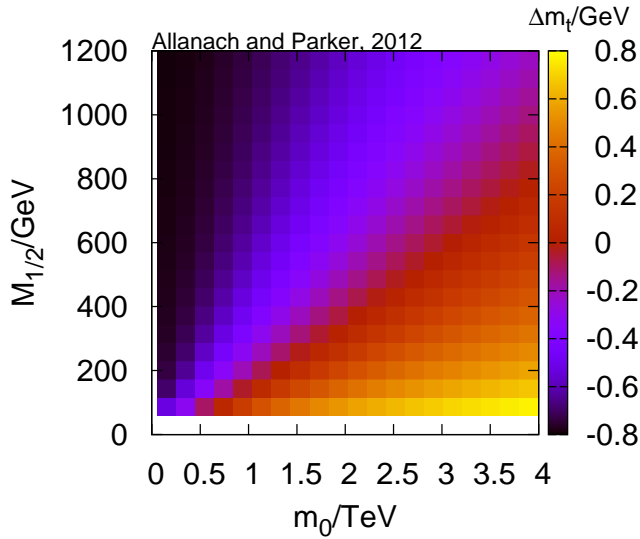}}
\end{picture}
\end{center}
\caption{\label{fig:dmt} Estimated size of higher order correction (2 loop
  SUSY QCD correction) 
  $(\Delta m_t)_{2LSQCD}(M_{SUSY})$ in the CMSSM for $\tan \beta=10$ and
  $A_0=0$. The   correction 
  has a size as measured by reference to the colour bar on the right hand
  side. }
\end{figure}

In our numerical analysis, we 
interpolate a grid of $m_0$ and $M_{1/2}$ 
values for $A_0=0$ and $m_b(m_b)=4.18$ GeV by using {\tt
  SOFTSUSY3.3.4}~\cite{Allanach:2001kg}.    
Fig.~\ref{fig:dmt} shows the estimated size of the higher order correction, as
determined in the way described above. Across the parameter space, we see that
$|(\Delta m_t)_{2LSQCD}(M_{SUSY})| < 0.8$ GeV. We may approximately model
the effect 
of this theoretical uncertainty by absorbing it into the pole mass
input uncertainty. If one wanted to do this more precisely, one would
renormalise 
$(\Delta m_t)_{2LSQCD}$ from $M_{SUSY}$ back to $M_Z$, where the running top
mass is matched to  the pole mass. However, the difference between including
$(\Delta m_t)_{2LSQCD}(M_{SUSY})$ and $(\Delta m_t)_{2LSQCD}(M_Z)$ is a
three-loop 
effect, which we neglect. It is possible that we have underestimated the
overall size of the theoretical uncertainty: after all, we have not included
possible effects from 3-loop RGEs~\cite{Ferreira:1996ug} or from 2-loop
threshold corrections to $m_t$~\cite{Bednyakov:2002sf}. 
We shall conservatively add the 0.8 GeV theoretical uncertainty {\em linearly}\/
to the 2$\sigma$ 
measurement errors upon the input value of $m_t$ in order to estimate the
total uncertainty in EWSB\@. 

\begin{figure}
  \begin{center}
\unitlength=1in
\begin{picture}(6,3)
  \put(-0.7,3){\includegraphics[angle=270,width=0.8\textwidth]{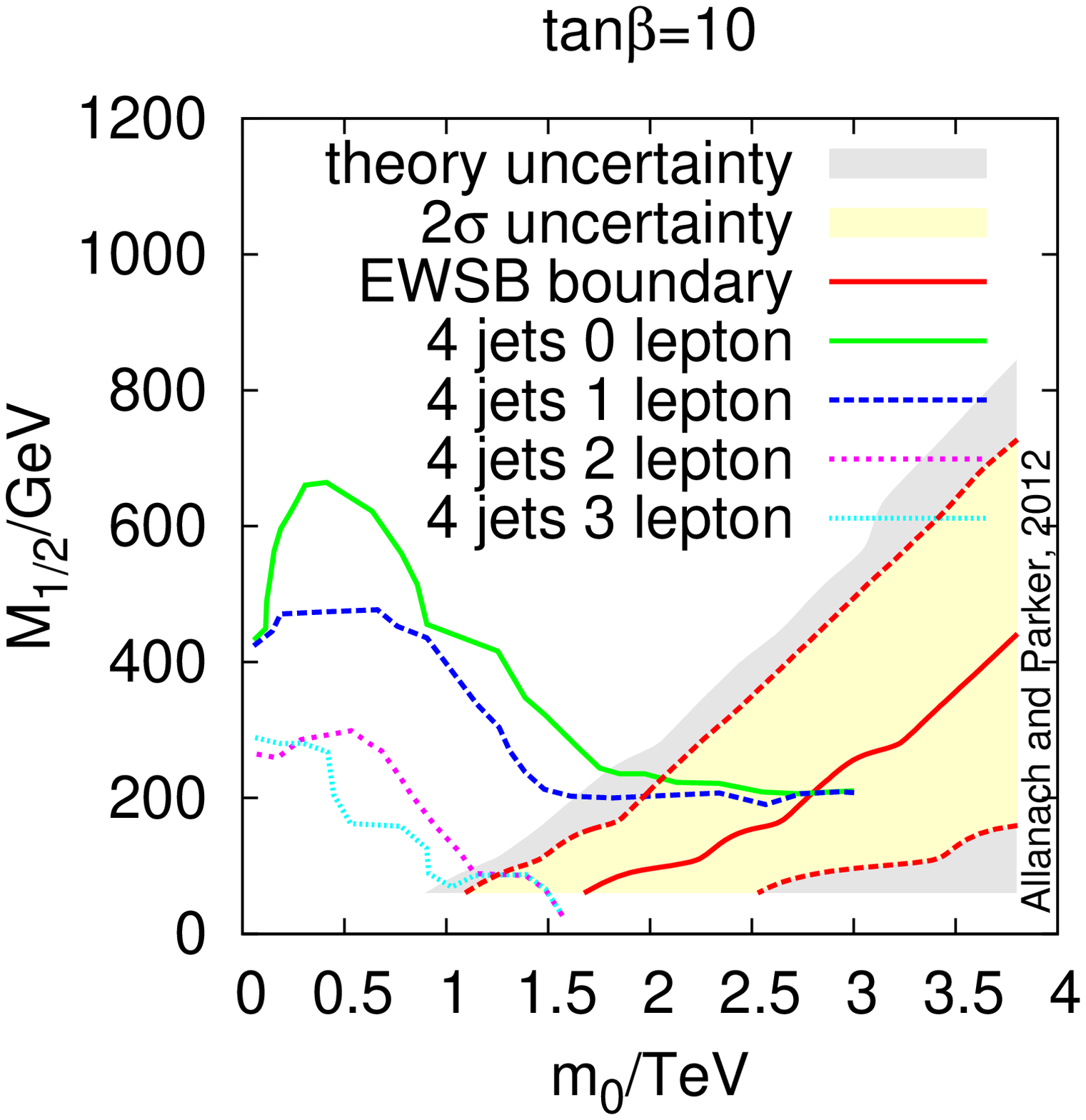}}
  \put(2.3,3){\includegraphics[angle=270,width=0.8\textwidth]{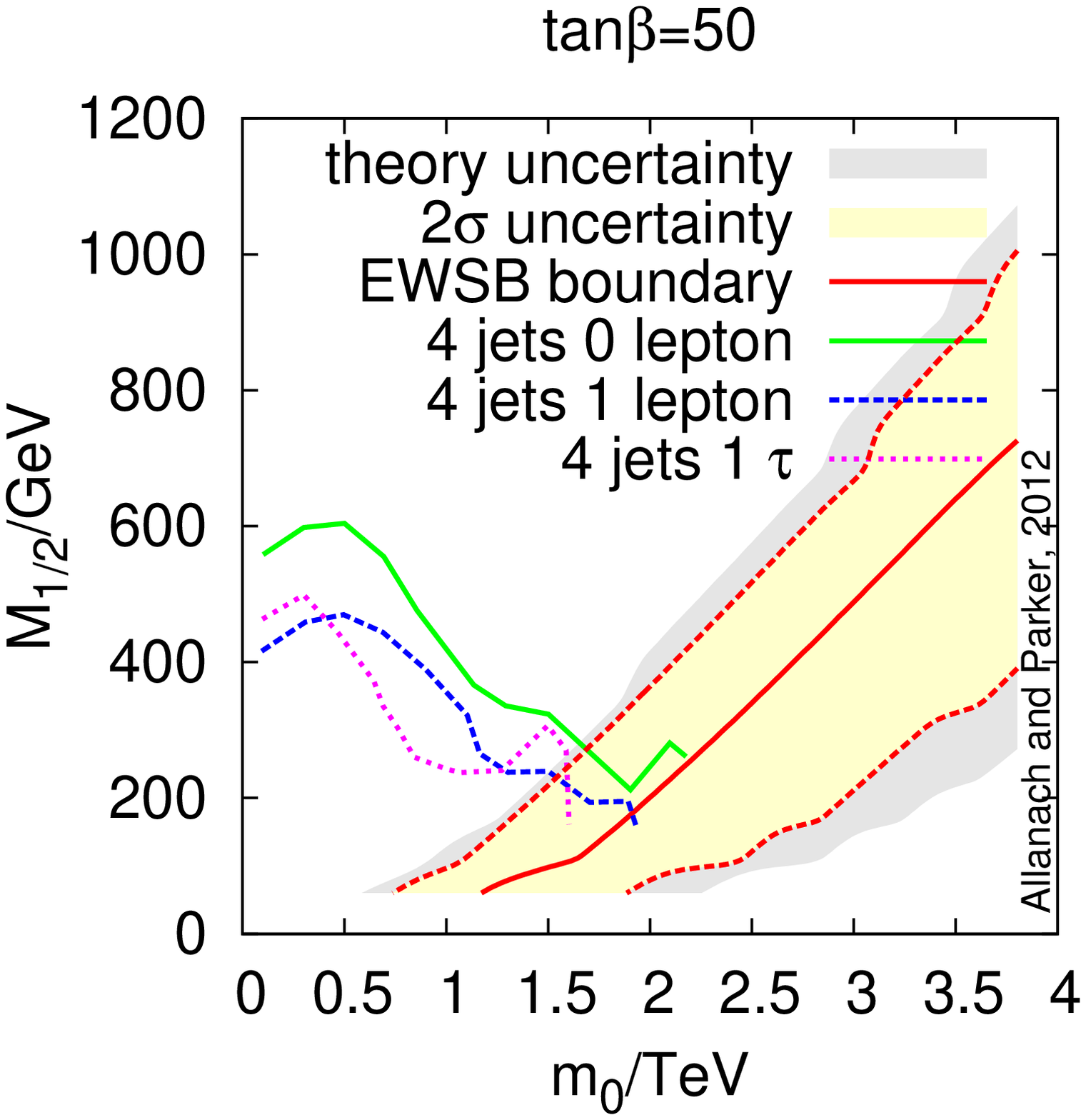}}
  \put(0.1,2.5){(a)}
  \put(3.1,2.5){(b)}
\end{picture}
  \end{center}
\caption{\label{fig:map} Uncertainties in the position of the EWSB
boundary compared to the expected 5$\sigma$ ATLAS 14 TeV CMSSM search
reach for 
1 fb$^{-1}$ from Ref.~\cite{Aad:2009wy}. The region below the `EWSB boundary' is
theoretically inaccessible: the uncertainty in the position of this region due
to 2$\sigma$ variations in $m_t$ is shown by the yellow band. The theoretical
uncertainty is marked by the grey region and 
estimated as detailed in the main body of the paper.}
\end{figure}

Having found and quantified the error due to the dominant theoretical
uncertainty, we may combine it with the $m_t$ measurement errors in order to
see the effect on the location of the EWSB boundary. 
The result is
shown in Fig.~\ref{fig:map} for $A_0=0$, (a) $\tan \beta=10$ and (b) $\tan
\beta=50$.  By varying the experimental uncertainty on
$m_t$ within 2$\sigma$ of its central value, we obtain the yellow (lighter)
shaded region showing the uncertainty in the boundary's location. The
curves marked `EWSB boundary' are, from left to right, for $m_t=171.5$ GeV,
$m_t=173.5$ GeV and $m_t=175.5$ GeV, respectively. Varying $m_t$ by a
further 0.8 GeV in either direction, we obtain the grey bands, providing an
estimate of the
theoretical uncertainty. 
We also show the expected 5$\sigma$ discovery reaches for 4 jets plus (0-3)
leptons for a 14 TeV LHC and 1 fb$^{-1}$ of luminosity from
Ref.~\cite{Aad:2009wy}.  
We see that the location of the boundary is uncertain to a huge $\sim 2$ TeV
in $m_0$, and that the discovery regions extend well into the region of
EWSB uncertainty, irrespective of the size of $\tan \beta$, even for only 1
fb$^{-1}$ of collected data. 
We also note that the 2$\sigma$ exclusion contours will be at much larger
values of $m_0$ and $M_{1/2}$, and so the problem of the uncertain EWSB
boundary will be more acute for exclusion than  for discovery. 
This combined with larger 14 TeV data sets implies that the EWSB uncertainty
is going to be very relevant for SUSY searches in the future. 

\begin{figure}
\unitlength=1in
\begin{center}
\begin{picture}(6,3)
  \put(-0.8,3){\includegraphics[angle=270,width=0.8\textwidth]{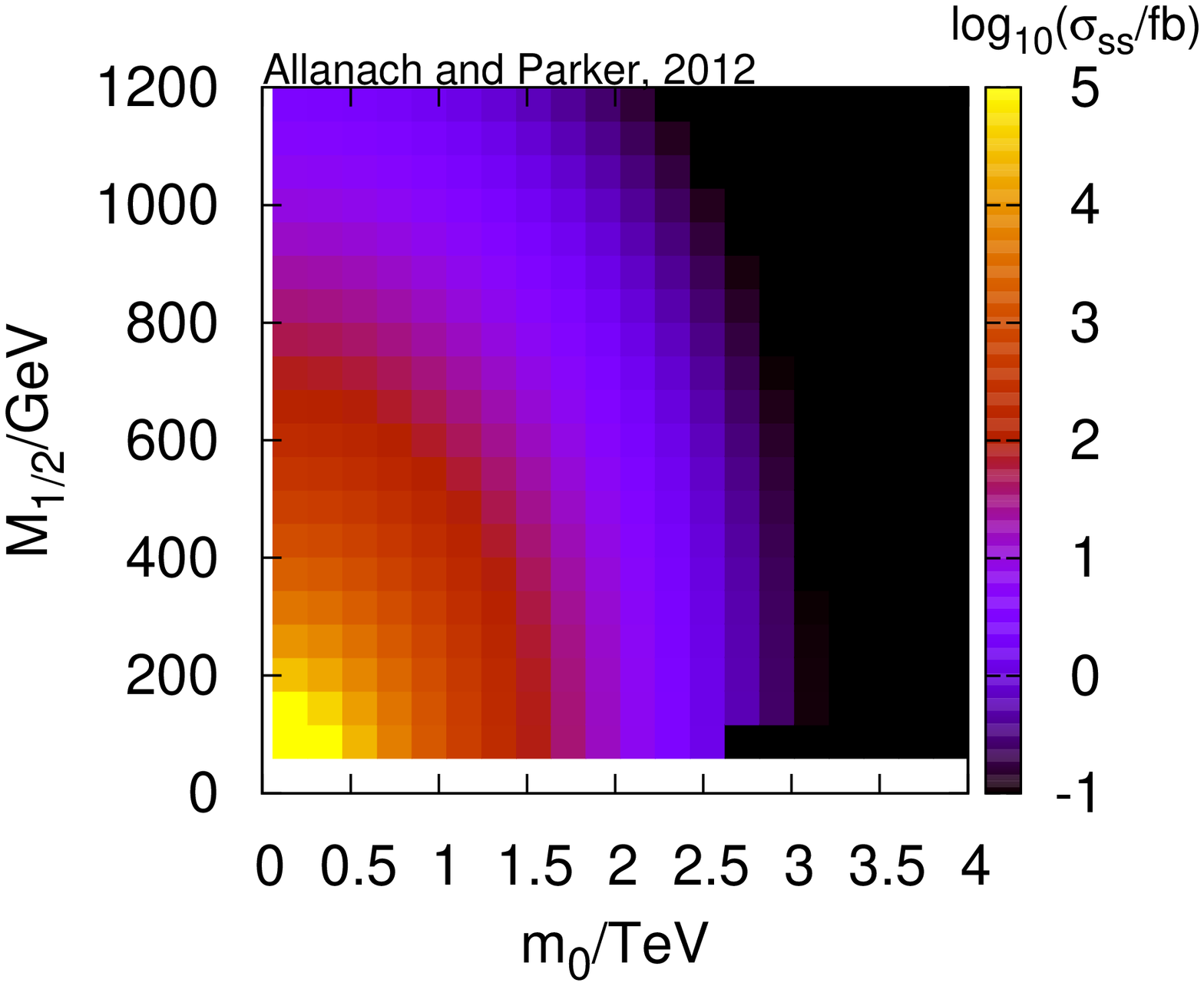}}
  \put(-0.01,2.54){\includegraphics[angle=270,width=0.54\textwidth]{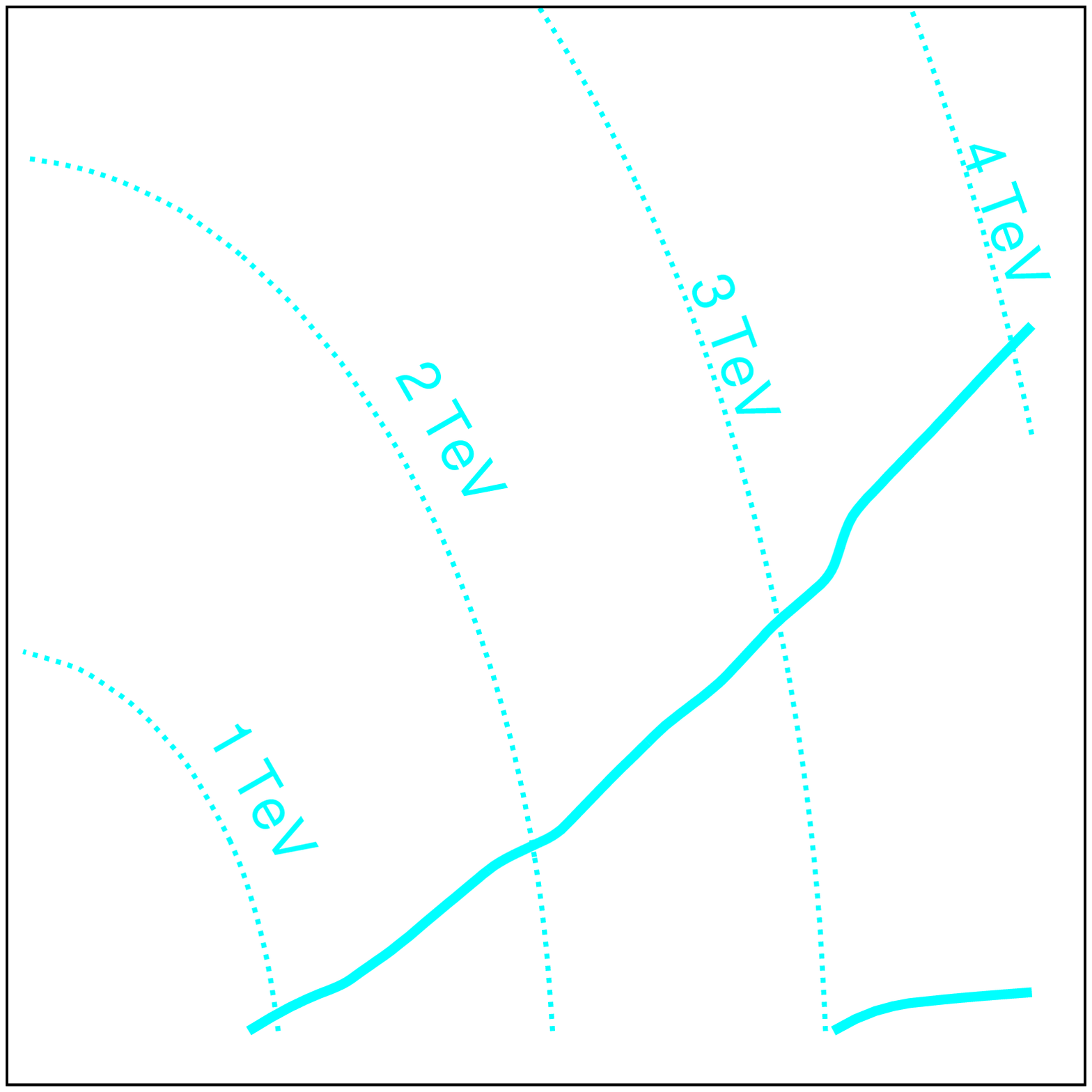}}
  \put(2.3,3){\includegraphics[angle=270,width=0.8\textwidth]{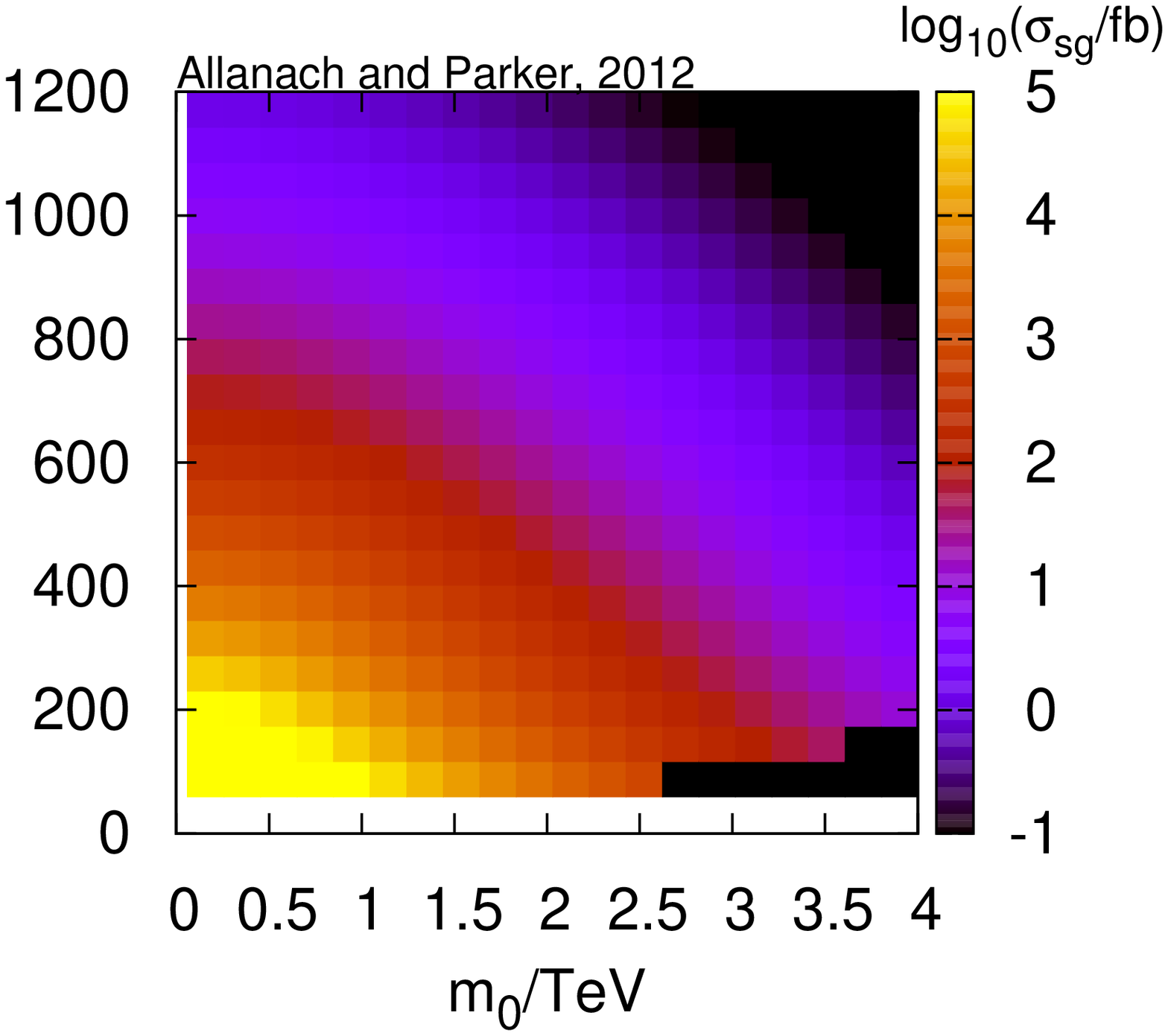}}
  \put(3.09,2.54){\includegraphics[angle=270,width=0.54\textwidth]{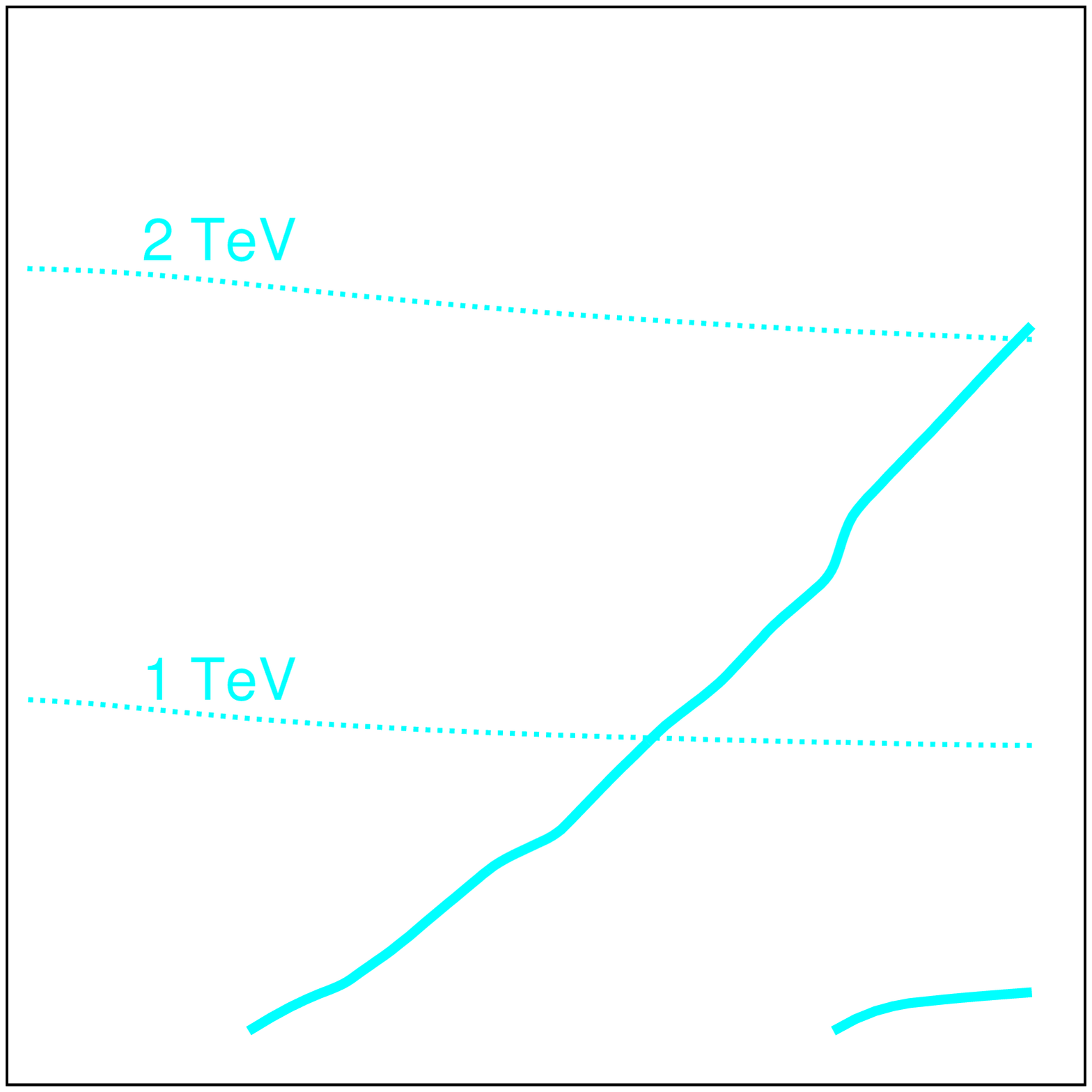}}
  \put(0,2.4){(a)}
  \put(3,2.4){(b)}
\end{picture}
\end{center}
\caption{\label{fig:sig} Uncertainties in 
  EWSB correlated with (a) squark-(anti-)squark and (b) squark
  gluino production cross-section at the 14 TeV LHC for $\tan \beta=10$ and
  $A_0=0$. The 
  two solid lines show the total
  uncertainty on the position of the boundary of EWSB\@. The total production
  cross-sections (with no cuts applied) are shown by reference to the colour
  bar on the right hand side. The dotted lines show (a) iso-contours of $m_{\tilde
    q}$ and (b) iso-contours of $m_{\tilde g}$, as labelled.}
\end{figure}

In Fig,~\ref{fig:sig}, we show that the size of the uncertainty can make a
large difference to the production cross-sections of squark (anti-)squark
$\sigma_{ss}$ and (anti-)squark gluino $\sigma_{sg}$ as calculated by {\tt
  PROSPINO2}~\cite{Beenakker:1996ch}, using the SUSY Les Houches
Accord~\cite{Skands:2003cj} to transfer information about the MSSM
spectrum. The solid lines show the  
limits of the uncertainty, as defined by both the experimental 2$\sigma$ $m_t$
uncertainties and the 0.8 GeV theoretical error acting in tandem. If, for
instance, we assume the CMSSM and constrain $M_{1/2}=400$ GeV from other
measurements, squarks may be highly  
visible at the LHC (when jointly produced with gluinos as shown
 in Fig.~\ref{fig:map}b) with large production
cross-sections (100 fb), or 
they may require a luminosity upgrade (1 fb or less) for discovery, depending
upon the locus of the boundary. Of course, this would not affect SUSY discovery
because discovery would be dominated by production of the lighter sparticles. 
However, the interpretation of supersymmetric signals in terms of a high-scale
SUSY breaking model
would be obscure unless the uncertainties are taken into account. For
example, it may be that the values of $m_0$ and $M_{1/2}$ inferred from
measurements of $\sigma_{sg}$ and other processes appear to be in the EWSB
region. One would then erroneously rule the CMSSM out.
When searches are interpreted near the boundary of EWSB therefore, in a
high-scale SUSY breaking model such as the CMSSM, it is
important to generate default SUSY spectra using a $m_t$ input value that is
2$\sigma$ 
{\em higher}\/ than the 
central value, so that the high end of the uncertain region is included in the
search. Increasing $m_t$ has a negligible effect on the gluino and first or
second generation squark
masses, so the production cross-sections should not change much. 
However, we show below that it has an effect
on the predicted lightest neutralino mass in the high-scale model. 
$m_t$ does have a larger effect on stop searches and so here, one
would need to check the effect's magnitude by taking the variation of the stop
masses from $m_t$ into account.

\begin{figure}
  \unitlength=1in
  \begin{center}
\begin{picture}(6,3)
  \put(-0.7,3){\includegraphics[angle=270,width=0.8\textwidth]{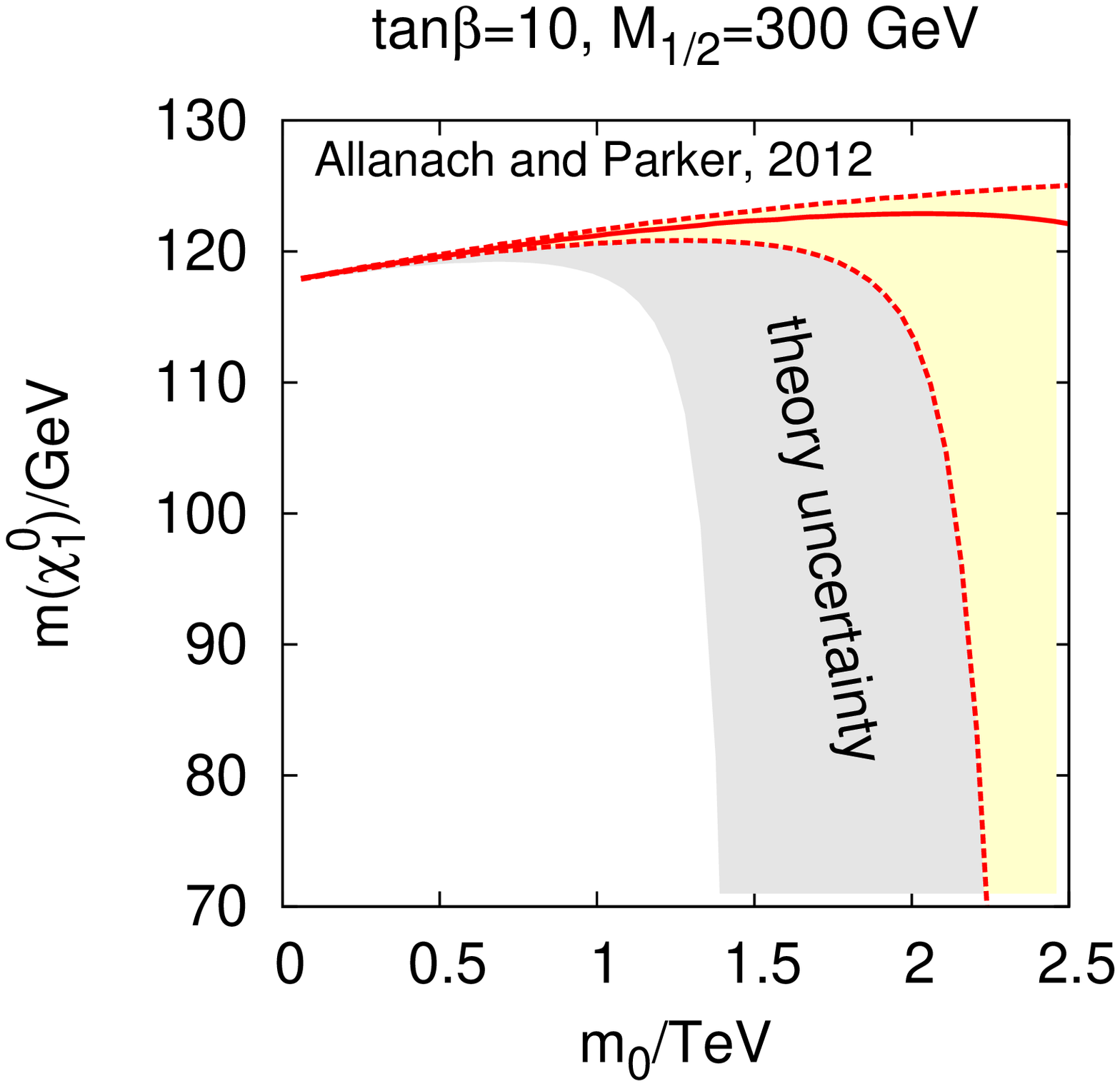}}
  \put(2.3,3){\includegraphics[angle=270,width=0.8\textwidth]{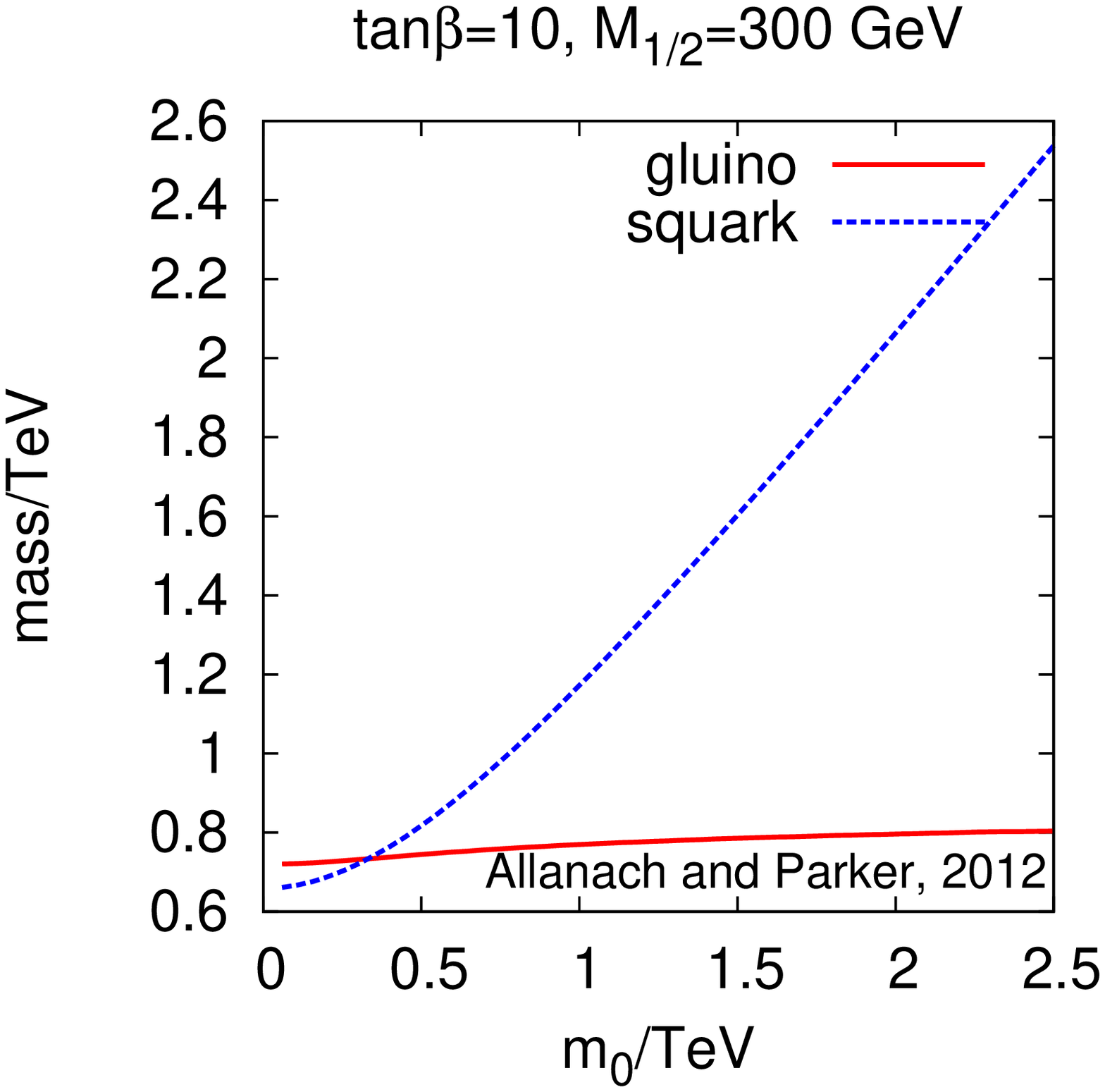}}
  \put(0.1,2.5){(a)}
  \put(3.1,2.5){(b)}
\end{picture}
\end{center}
\caption{\label{fig:dmchi} Evolution in certain SUSY particle masses
  approaching the EWSB boundary for $\tan 
  \beta=10, M_{1/2}=300$ GeV and $A_0=0$ in the CMSSM\@.
(a) shows uncertainties in the CMSSM predicted lightest neutralino mass.
   The solid line has $m_t=173.5$ GeV, whereas
  the upper and lower dashed lines have $m_t=175.5$ GeV and $m_t=171.5$ GeV,
  respectively. Additional uncertainty coming from the higher order
  theoretical uncertainty in $m_t$
  via RGE effects is marked by the grey
  region. (b) shows values of the squark and gluino masses as a function of
  $m_0$ along the $M_{1/2}=300$ GeV line. } 
\end{figure}

On the other
hand, we know that $\mu \rightarrow 0$ at the EWSB boundary, 
affecting the mass of the lightest supersymmetric particle (LSP), here assumed
to be the lightest neutralino. 
The MSSM Lagrangian contains the neutralino mass matrix as
$-\frac{1}{2}
{\tilde\psi^0}{}^T{\cal M}_{\tilde\psi^0}\tilde\psi^0$ + h.c., where
$\tilde\psi^0 =$ $(-i\tilde b,$ $-i\tilde w^3,$ $\tilde h_1,$ $\tilde
h_2)^T$ and, at tree level,
\begin{equation}
{\cal M}_{\tilde\psi^0} \ =\ \left(\begin{array}{cccc} M_1 & 0 &
-M_Zc_\beta s_W & M_Zs_\beta s_W \\ 0 & M_2 & M_Zc_\beta c_W &
-M_Zs_\beta c_W \\ -M_Zc_\beta s_W & M_Zc_\beta c_W & 0 & -\mu \\
M_Zs_\beta s_W & -M_Zs_\beta c_W & -\mu & 0
\end{array} \right). \label{mchi0}
\end{equation}
We use $s$ and $c$ for sine and cosine, so that
$s_\beta\equiv\sin\beta,\ c_{\beta}\equiv\cos\beta$ and $s_W (c_W)$ is
the sine (cosine) of the weak mixing angle.  The condition
$\mu \rightarrow 0$ (at the EWSB boundary) results in a lightest
neutralino mass (equal to the smallest eigenvalue of Eq.~\ref{mchi0})
that is zero at tree-level\footnote{One can easily see this by 
the observation that adding $\tan \beta$ multiplied by the third column to
the fourth column leads to a column of zeroes in the $\mu \rightarrow 0$ limit. The resulting matrix, which has
a determinant of $\tan \beta$ multiplied by the determinant of the original
matrix, must therefore have a zero determinant and therefore possesses a zero
eigenvalue.}. Including loop corrections 
to the neutralino mass 
matrix, the neutralino LSP is still very
light, typically less than 1 GeV.  
As the position of the EWSB boundary is so uncertain, 
the position of this negligible neutralino mass occurs for
very different values of $m_0$. 
Since $m_0$ controls the squark
mass, the correlation between squark and LSP mass
picks up large uncertainties. We illustrate the problem by scanning along
the $M_{1/2}=300$ GeV line for various different values of $m_t$, including an
estimate of the theoretical uncertainty originating from higher order
corrections to $m_t$ as described above. From Fig.~\ref{fig:dmchi}a, we see
the large uncertainty induced in the lightest neutralino mass along the line. 
We see from Fig.~\ref{fig:dmchi}b that the high $m_0$ region where the neutralino mass is particularly
uncertain corresponds to large values of the squark masses $>1.5$ TeV. Here,
the classic jets plus 
missing transverse momentum signature will come primarily from $\tilde g
\tilde g$ production, where $\tilde g$ decays through an off-shell squark into 
two jets and the neutralino LSP\@. In the centre of mass frame of the 700-800
GeV gluino, 
the gluino must share its rest mass energy between the two jets and the lightest
neutralino.  Having higher neutralino masses therefore has the effect of
reducing the $p_T$ of the jets, on average. Since experimental analyses
impose lower 
limits on the jet $p_T$s in the cuts, the efficiency of the cuts will
decrease for higher neutralino masses. We therefore expect the
theoretical uncertainty in the neutralino mass to induce a non-negligible
uncertainty in the cut efficiency. 

\begin{figure}
\unitlength=1.15in
\begin{center}
%\begin{picture}(3,3.5)
\begin{picture}(6,2.2)
\put(0.5,0) {\includegraphics[angle=0,width=0.5\textwidth]{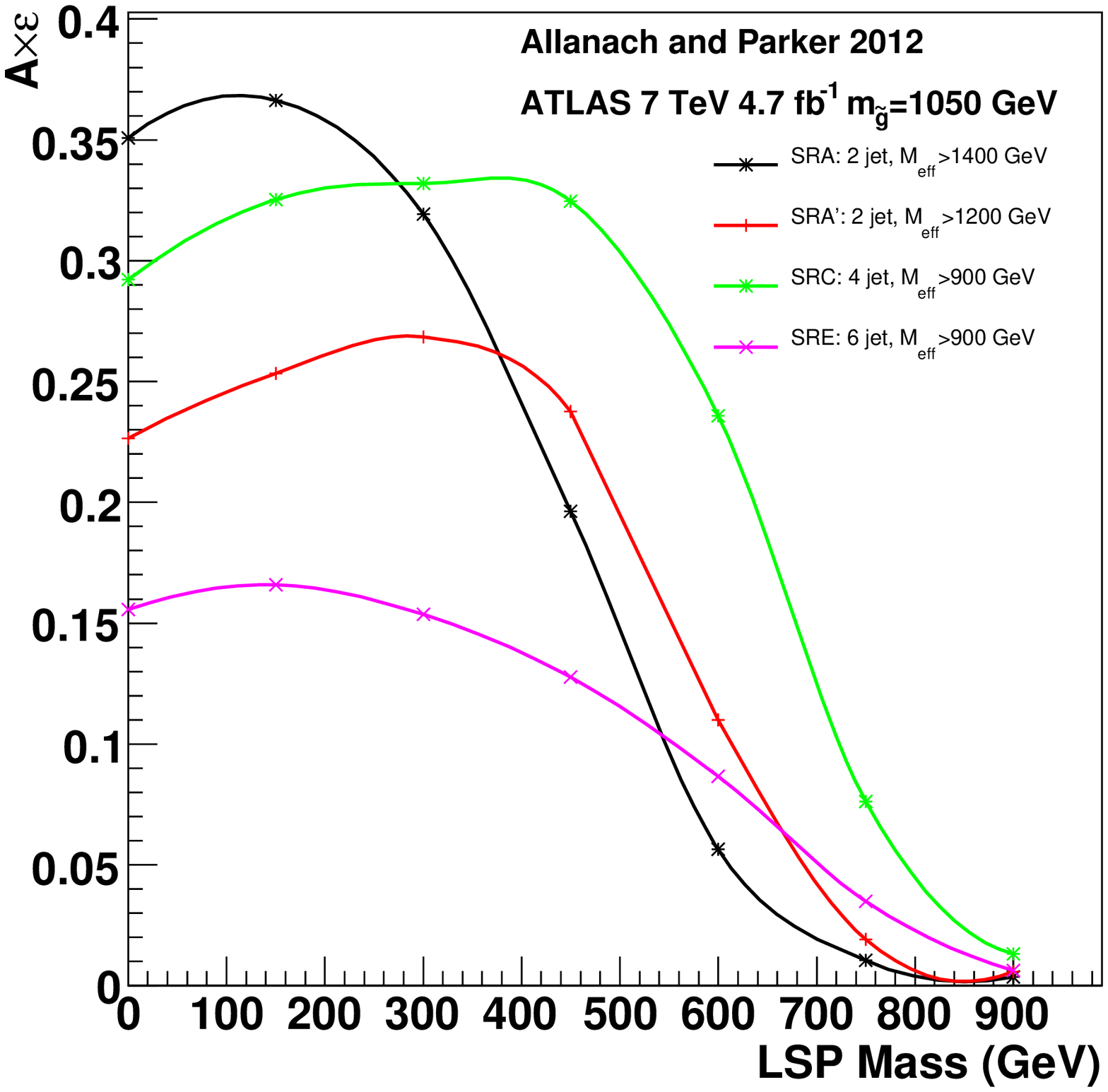}}
\put(3.3,0){\includegraphics[angle=0,width=0.5\textwidth]{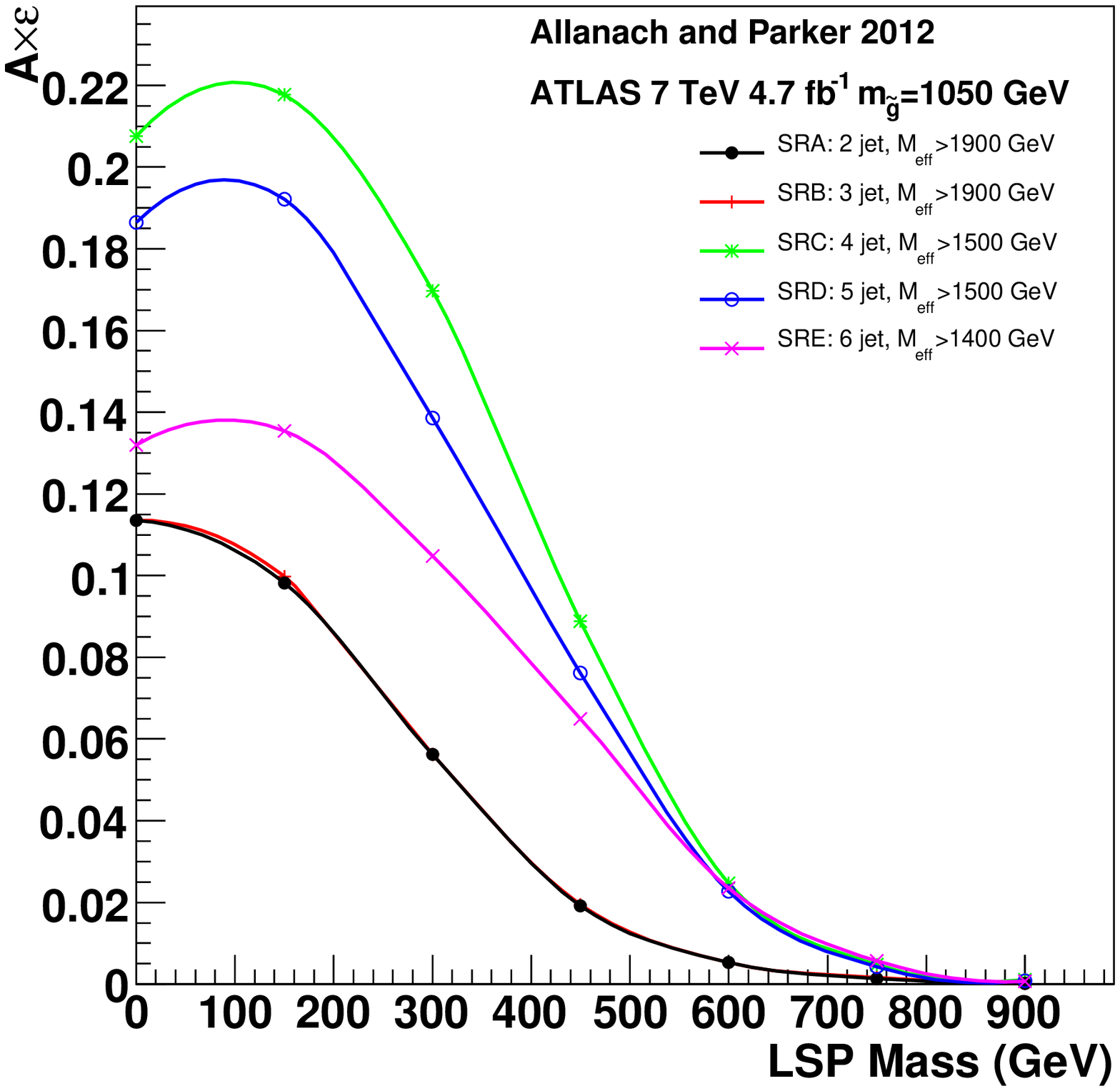}}
  \put(-0.1,2){(a)}
  \put(2.7,2){(b)}
\end{picture}
\end{center}
\caption{\label{fig:atimese} Acceptance times efficiency $A \times
  \epsilon$ for signal events
  selected by the ATLAS supersymmetry search \cite{Aad:2012rz}, using the
  signature of missing energy with hadronic jets. The different signal regions
  correspond to final state jet multiplicities (n) between 2 and 6, as shown
  in the key. The signal regions are selected with minimum values of
  $m_{\mathrm{eff}}$, the scalar sum of the $p_T$s of the n leading jets
  (appropriate to each signal region) with the missing energy. The lowest
  available $m_{\mathrm{eff}}$ selection is shown in (a), and the highest in (b)
  for various jet multiplicities.} 
\end{figure}

In Figure~\ref{fig:atimese} we plot data on the acceptance times efficiency
($A\times \varepsilon$) for the ATLAS search based on all hadronic final states
with missing energy \cite{Aad:2012rz}. These data, for the 7 TeV sample, are
available in the auxiliary information of Ref.~\cite{Aad:2012rz}. The data
correspond to a 
model in which the gluino decays directly to jets and LSP, with the gluino
mass fixed at 1050 GeV, close to the EWSB boundary. The other SUSY particle masses are set high enough for them to decouple. The figures show the
fraction of signal events which would pass the experimental selections
(based largely on the $p_T$ of the jets, and the missing transverse momentum,
and also be 
successfully reconstructed after detection, as a function of the LSP
mass. Figures~\ref{fig:atimese}a and b show the lowest and highest effective
mass selections for each jet multiplicity, respectively. The 
value of $A\times \varepsilon$ drops rapidly at high LSP mass for all of the
signal regions. This is expected, 
since the fraction of the initial squark or gluino energy passing into the
observable jets is reduced as the LSP mass rises, and hence the fraction
passing the $p_T$ selections drops. For the lowest multiplicity final states
(SRA), with the higher effective mass selection, the effect can be seen even
at the lowest LSP masses, with $A\times 
\varepsilon$ reducing from 11.7\% to 9.8\% as the LSP mass rises from 0 to 150
GeV. This represents a change in signal event rate of 17\%, a non-negligible
uncertainty. In the lower effective mass samples, $A\times\varepsilon$ rises
in the same LSP mass range in some cases (an effect already noted in
Ref.~\cite{LeCompte:2011fh}). For example, for SRA', it increases
from  
23.5\% to 26.1\%, increasing the signal rate by  11\%, and continues to rise
for LSP masses up to 300 GeV. This may be due to the behaviour of the cut on the ratio of the missing energy to $m_{\mathrm{eff}}$ applied in this analysis, since the two distributions will evolve in slightly different ways as the LSP mass changes. 
These effects can be expected to continue to occur at higher centre of mass
energies, with the final drop in $A\times \epsilon$ moving to higher LSP
masses.  

Even if the high-scale SUSY breaking
interpretation of an experiment's sparticle search is largely
independent of the uncertainty induced in the LSP mass, it is
possible that other sparticle masses in the signal sparticle cascade decay
chains {\em are}\/ highly dependent and cause large theoretical uncertainties in
signal cut 
efficiencies. In particular, 
chargino masses are highly dependent on $\mu(M_{SUSY})$. Also, third
generation sparticles may be highly dependent on $\mu(M_{SUSY})$ in some
(but not all) regions of parameter space. Extra care to quantify the
theoretical error should be taken in
searches for cascade chains
involving these sparticles, particularly when they are on-shell (since the
mass splittings of on-shell decays may have a particularly large effect on the
kinematics of the decay products). 
In order to take such uncertainties into account,
varying $m_t$ within its 2$\sigma$ limits, including the size of the shift
from higher order corrections (estimated here to be 0.8 GeV) 
in the estimate of the uncertainty will suffice, since the various relevant
sparticle masses will vary
according to the concomitant theoretical error. 

\section{Convergence \label{sec:con}}

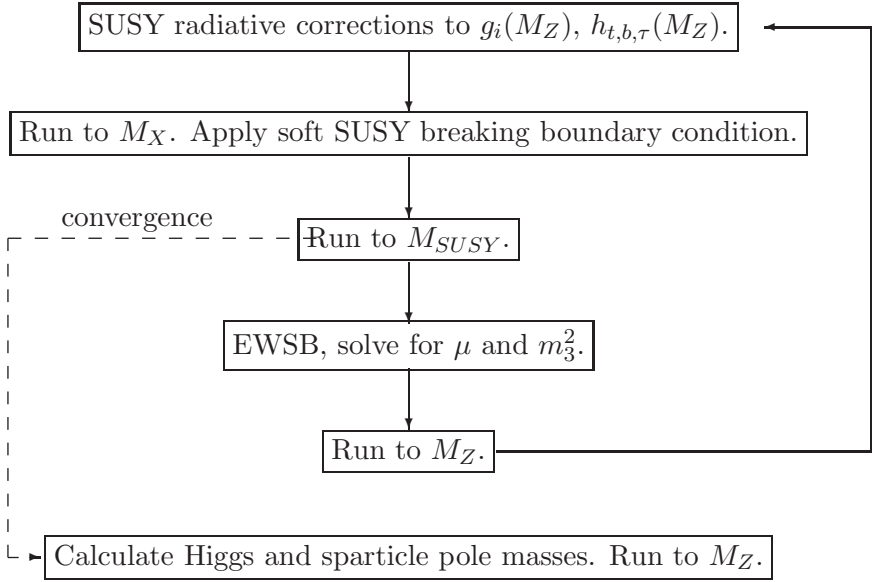
\begin{figure}\begin{center}
\label{fig:algorithm}
\begin{picture}(323,210)
\put(10,0){\makebox(280,10)[c]{\fbox{Calculate Higgs and
      sparticle pole masses. Run to $M_Z$.}}}
%\put(150,36.5){\vector(0,-1){23}}
\put(10,40){\makebox(280,10)[c]{\fbox{Run to $M_Z$.}}}
\put(150,76.5){\vector(0,-1){23}}
\put(10,160){\makebox(280,10)[c]{\fbox{Run to $M_X$. Apply soft SUSY breaking
boundary condition.}}}
\put(150,116.5){\vector(0,-1){23}}
\put(10,80){\makebox(280,10)[c]{\fbox{EWSB, solve for $\mu$ and $m_3^2$.}}}
\put(150,156){\vector(0,-1){23}}
\put(10,120){\makebox(280,10)[c]{\fbox{Run to $M_{SUSY}$.}}}
\DashLine(115,125)(0,125){5}
\DashLine(0,125)(0,5){5}
\DashLine(0,5)(12,5){5}
\put(10,5){\vector(1,0){2}}
\put(20,130){convergence}
\put(150,196){\vector(0,-1){23}}
\put(10,200){\makebox(280,10)[c]{\fbox{SUSY radiative corrections to
$g_i(M_Z)$, $h_{t,b,\tau}(M_Z)$.}}}
\put(183,45){\line(1,0){140}}
\put(323,45){\line(0,1){160}}
\put(323,205){\vector(-1,0){40}}
\end{picture}
\caption{Iterative algorithm used to calculate the SUSY spectrum. 
The initial step is the
uppermost one. $M_{SUSY}$ is the scale at which the EWSB
conditions 
are imposed, as discussed in the text. $M_X$ is the scale at which the high
energy SUSY breaking boundary conditions are imposed.}\end{center}\end{figure}
\begin{figure}
  \begin{center}
    \includegraphics[angle=270,width=0.8\textwidth]{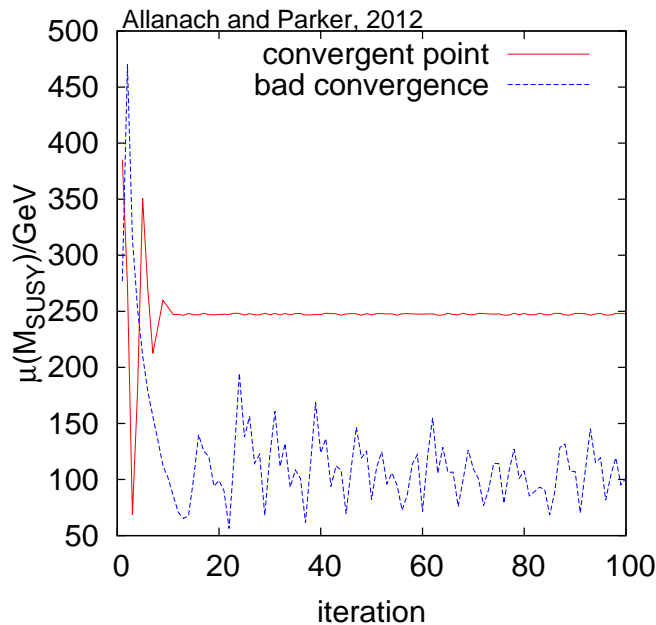}
  \end{center}
\caption{\label{fig:points} Convergence of the numerical solution to
  EWSB Higgs potential minimisation conditions near
  the boundary of electroweak symmetry breaking in the CMSSM using {\tt
    SOFTSUSY3.3.4}. We have set 
  $\epsilon=10^{-3}$,   $m_t=173.5$ GeV, 
  $\tan \beta=10$, $A_0=0$, $M_{1/2}=337.5$ GeV and $m_0=3000$ GeV for the
  `convergent point', or $m_0=$3400 GeV for the `bad convergence' point.}
\end{figure}

As one approaches the boundary of EWSB in the
parameter space of high-scale supersymmetry breaking models, one finds that the
various spectrum generators start to struggle to produce numerically stable
results. This is because of the extreme sensitivity of the RGEs and boundary
conditions, but is not a problem if one is only fixing the SUSY breaking
parameters at the weak scale. {\tt SOFTSUSY3.3.4} is essentially solving
coupled, non-linear 
ordinary differential equations with two sets of boundary conditions: one set
on the gauge couplings $g_i$, the Yukawa couplings $h_{t,b,\tau}$ 
and the Higgs potential parameters $\mu$ and $m_3^2$ at the weak scale, and
the other set on the soft SUSY breaking parameters at the high scale $M_X$.
{\tt SOFTSUSY3.3.4} uses an iterative algorithm, as shown in
Fig.~\ref{fig:algorithm}. For {\tt SOFTSUSY3.3.4} to report that it has solved
the problem, a convergence criterion must be satisfied: the values of the
$\overline{DR}$ sparticle masses of the current iteration must all be
identical to the previous iteration's, to some fractional accuracy $\epsilon$. 
Here, $\epsilon=10^{-3}$. Near the boundary of EWSB, 
$\mu(M_{SUSY})$ and $m_3(M_{SUSY})$ become extremely dependent upon the
precise value of $h_t(M_{SUSY})$. However, $h_t(M_{SUSY})$, which is
dependent on $m_t(M_{SUSY})$, varies from iteration to iteration because 
the loop corrections depend upon the sparticle masses, which also vary from
iteration to iteration. In parameter space that is sufficiently near to the
boundary of EWSB, this 
numerical iteration noise becomes too great for the program to find a
solution, as illustrated in Fig.~\ref{fig:points}. We show two
points in CMSSM parameter space, one with bad convergence, and one with good
convergence. We display the predicted value of $\mu(M_{SUSY})$, which is one
of EWSB variables that is most sensitive. We see that after 10 iterations, the
convergent point has found a stable solution for $\mu(M_{SUSY})$, whereas
there is no indication that the point with bad convergence would ever
converge. In practice, if no convergence is achieved after 30 iterations, {\tt
  SOFTSUSY3.3.4} halts the calculation reporting a `no convergence' error. 

\begin{figure}
\unitlength=1in
\begin{center}
\begin{picture}(6,2.5)
  \put(-0.7,3){\includegraphics[angle=270,width=0.8\textwidth]{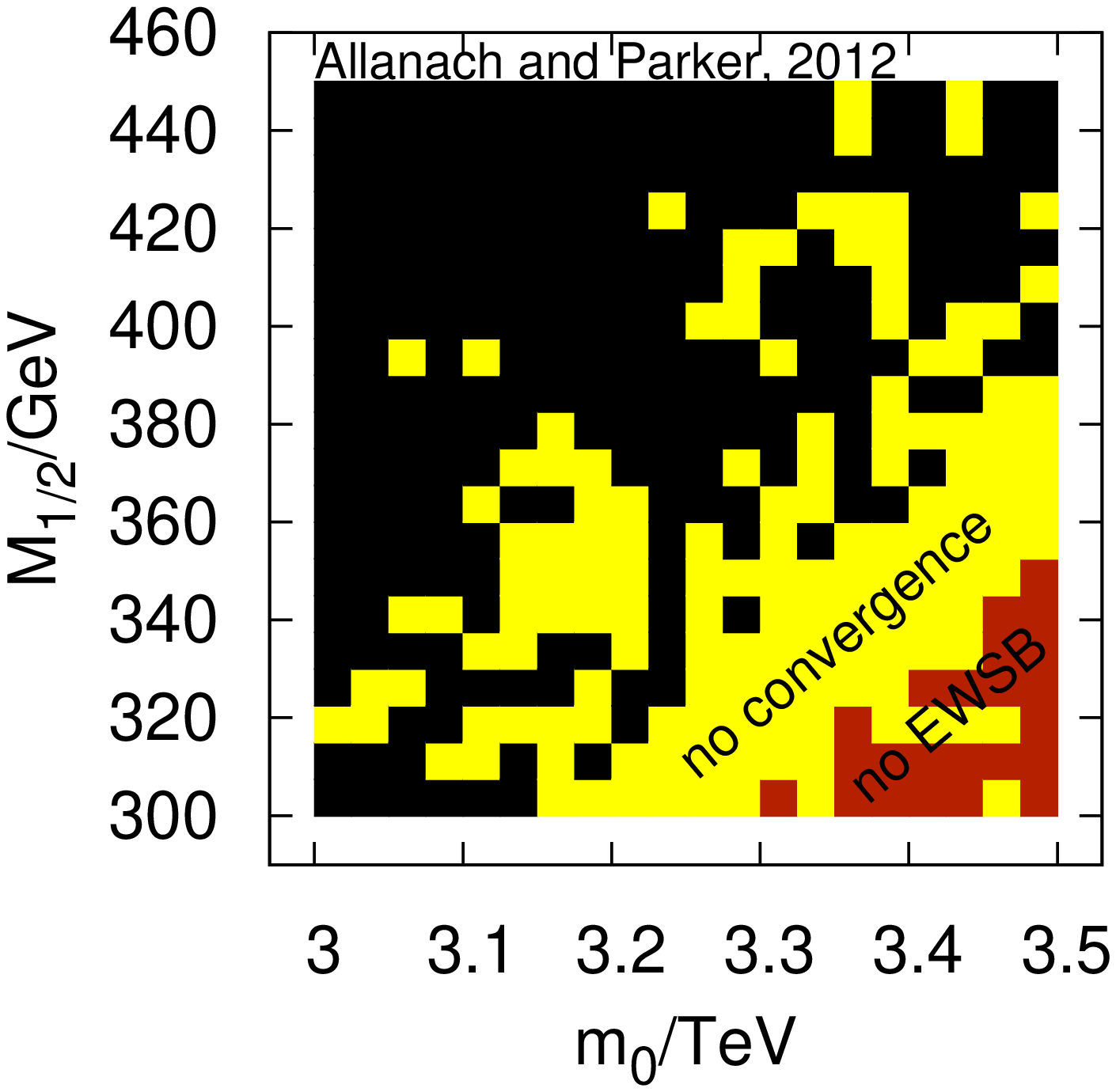}}
  \put(0.09,2.54){\includegraphics[angle=270,width=0.54\textwidth]{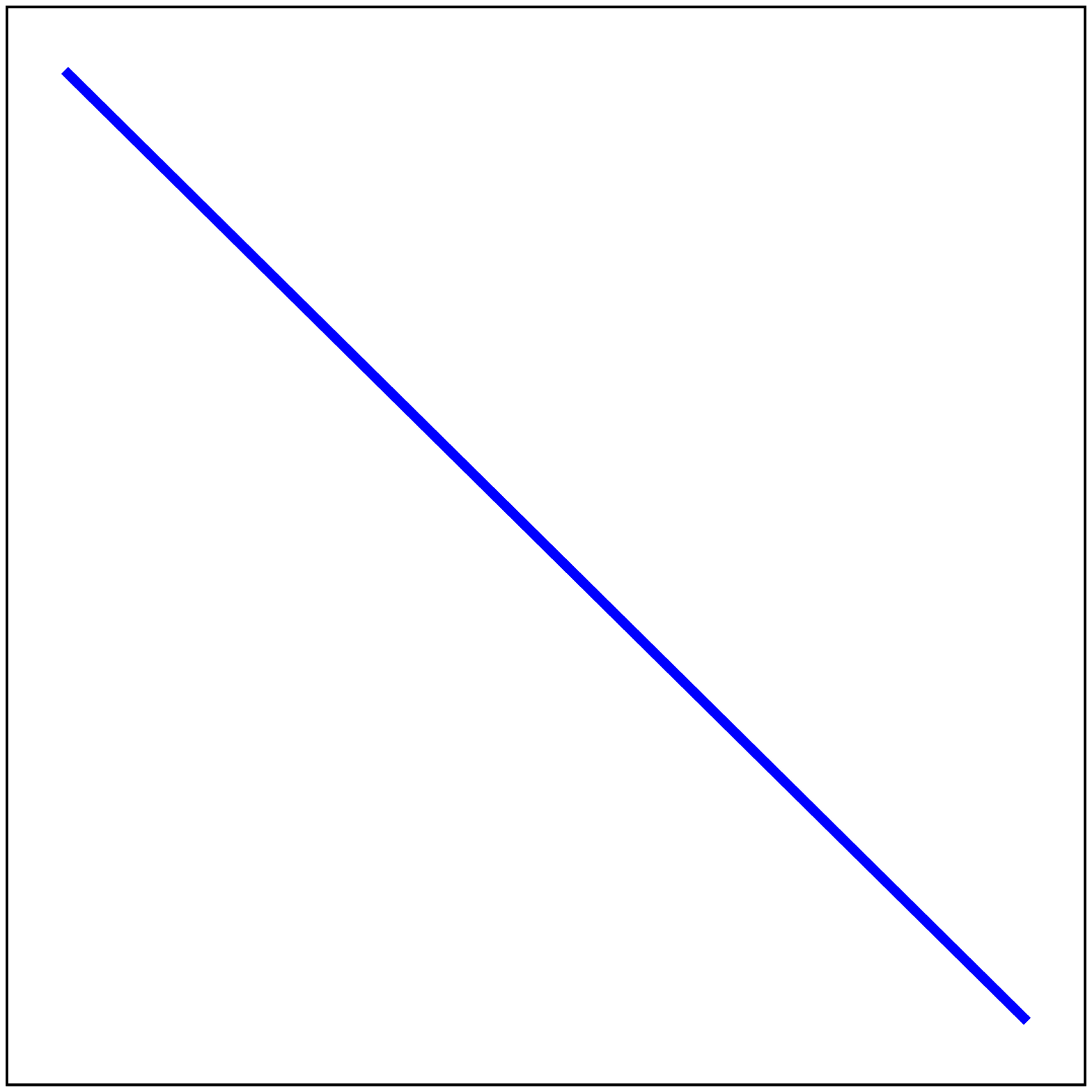}}
  \put(2.3,3){\includegraphics[angle=270,width=0.8\textwidth]{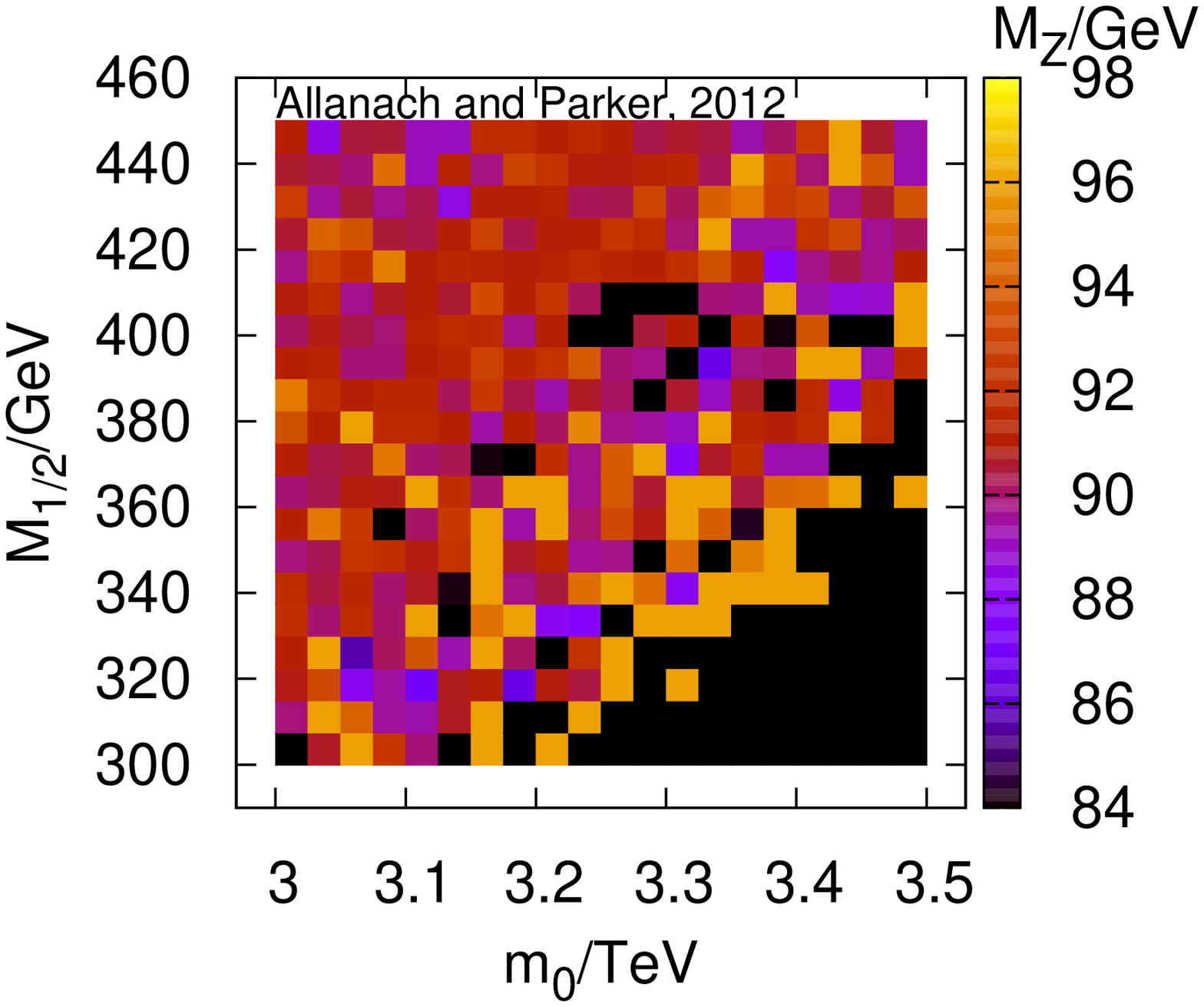}}
%  \put(3.09,2.54){\includegraphics[angle=270,width=0.54\textwidth]{line}}
  \put(0,2.4){(a)}
  \put(3,2.4){(b)}
\end{picture}
\caption{\label{fig:scan} Zoom of CMSSM focus point parameter space as
  predicted by 
{\tt SOFTSUSY3.3.4} for $A_0=0$, $m_t=173.5$ GeV and $\tan \beta=10$: (a)
shows the 
  region of successful EWSB (black), bad numerical
  convergence (yellow) and 
no EWSB (red). The
diagonal line shows the parameter line 
taken to illustrate non-convergence in Fig.~\protect\ref{fig:line}. (b) displays the predicted value of
$M_Z/GeV$, as measured by the colour bar to the right hand side.}\end{center}
\end{figure}

\begin{figure}
  \begin{center}
    \includegraphics[angle=270,width=0.7\textwidth]{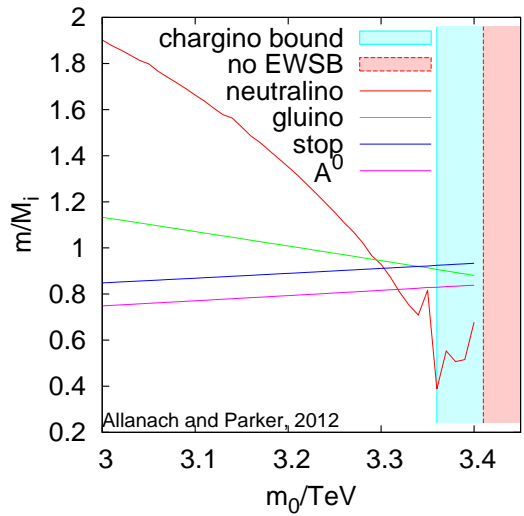}
\caption{\label{fig:line} {\tt SOFTSUSY3.3.4} convergence approaching the
  boundary of EWSB in the CMSSM for $m_t=173.5$ GeV. The plot shows the blue
  line in 
  Fig.~\protect\ref{fig:scan} (i.e.\ $M_{1/2}$ is varying across the plot as
  well as $m_0$). We have scaled each $i^{th}$ particle mass by a constant mass
  $M_i$ for clarity. The region where electroweak symmetry is not
  satisfactorily broken is denoted `no EWSB'. $m_{{\chi}^\pm_1}>104$ GeV, in
  contravention of the LEP bound, in the area marked
  `chargino bound'.}  \end{center}
\end{figure}

If one's scans in parameter space are sufficiently coarse, one may not be
aware of this problem, as the region of no convergence is fairly thin in the
$m_0$-$M_{1/2}$ plane. We zoom in on the plane in Fig.~\ref{fig:scan} in
order to investigate it. Fig.~\ref{fig:scan}a shows that the yellow (light)
regions of bad convergence are prevalent close to the boundary of no EWSB\@. Using
Eq.~\ref{higgsMin} to predict $M_Z$ for the numerical solution that
issues 
from {\tt SOFTSUSY3.3.4} (whether or not convergence is reached), we plot its
value in the zoomed region in Fig.~\ref{fig:scan}b. We see that throughout
this zoomed region, despite the fact that central experimental value of
$M_Z$ 
has been used at each iteration to constrain the model, the large sensitivity
spoils the prediction after each iteration and the predicted value is far from
the central one of $M_Z^{cent}=91.1876$ GeV~\cite{Beringer:1900zz}. The
diagonal line in Fig.~\ref{fig:scan}a 
across parameter space shows the line taken in Fig.~\ref{fig:line} to
illustrate how 
how the numerical spectral predictions of {\tt SOFTSUSY3.3.4} vary while
travelling 
through this problematic region of parameter space. Each line represents a
different sparticle mass divided by some constant number of GeV, chosen so
that the ratios are all similar and appear near to each other on the same plot. 
We see that the gluino,
stop and pseudo-scalar Higgs masses have a smooth behaviour as the region of no
EWSB is approached. However, we see that the neutralino LSP mass, which
depends upon $\mu(M_{SUSY})$, becomes erratic close to the boundary,
corresponding to numerical problems and non-convergence of the iteration
algorithm. Using the numerical solution at the inferred boundary but
substituting $\mu=0$ into the one loop level neutralino mass matrix, we infer
a mass of 0.1 
GeV for the neutralino. The neutralino line would be at $m/M \approx 0.01$ at
this 
point, rather than around 0.6.  
Some of the erratic non-convergent behaviour can be hidden behind the LEP
chargino bound of $M_{\chi^\pm}>104$ GeV~\cite{Nakamura:1299148}, but still
some noise remains unhidden. 

\begin{figure}
\unitlength=1in
\begin{center}
\begin{picture}(6,2.5)
  \put(-0.7,3){\includegraphics[angle=270,width=0.8\textwidth]{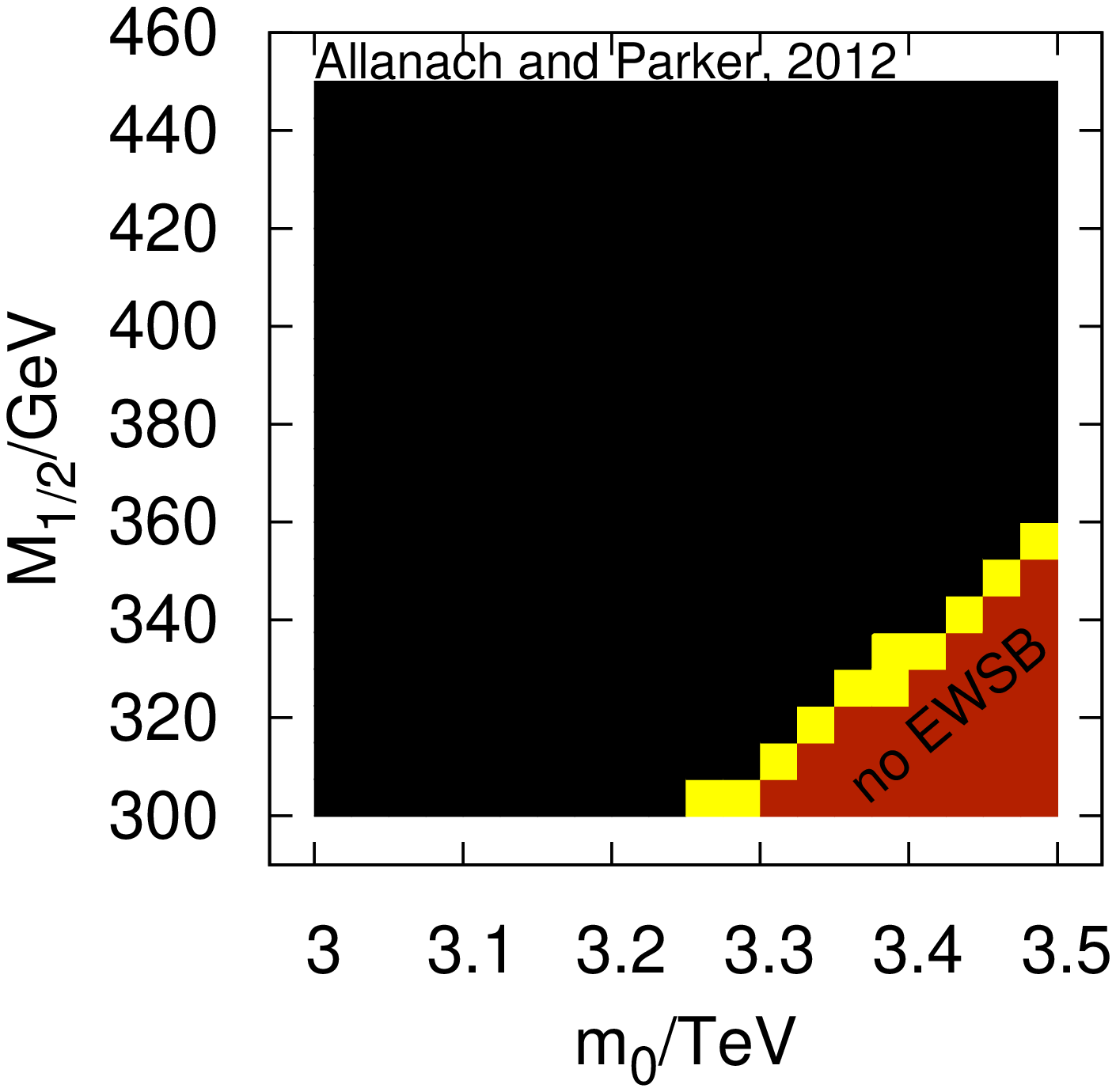}}
%  \put(0.11,2.54){\includegraphics[angle=270,width=0.54\textwidth]{mchiBd}}
  \put(2.3,3){\includegraphics[angle=270,width=0.8\textwidth]{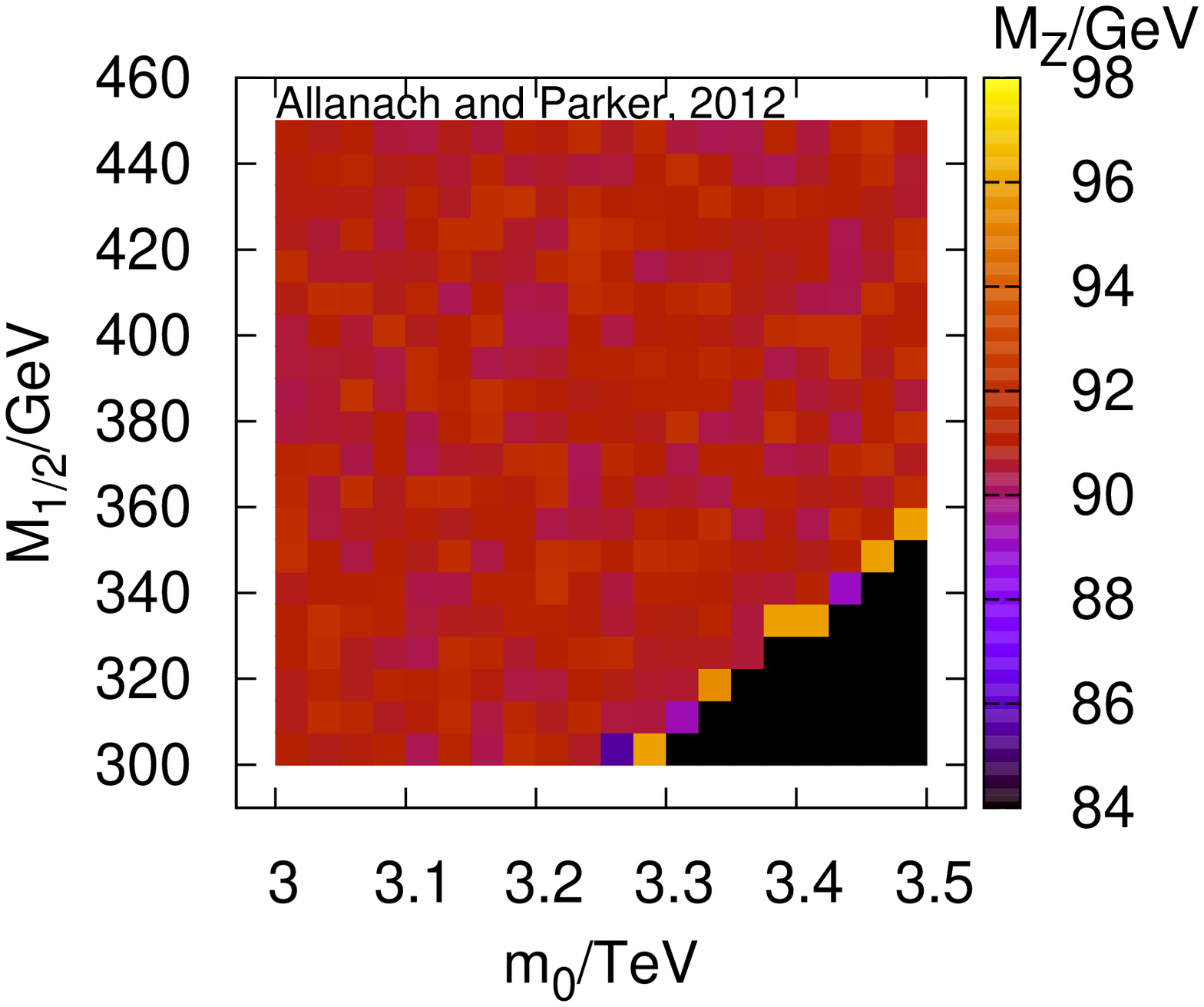}}
%  \put(3.09,2.54){\includegraphics[angle=270,width=0.54\textwidth]{line}}
  \put(0,2.4){(a)}
  \put(3,2.4){(b)}
\end{picture}
\caption{\label{fig:scanImp} Zoom of CMSSM focus point parameter space with the
  improved algorithm in {\tt SOFTSUSY 3.3.6}
 for $m_t=173.5$ GeV: (a) shows the
  region of successful EWSB (black), bad numerical
  convergence (yellow) and 
no EWSB (red). (b) displays the predicted value of
$M_Z$/GeV, as measured by the colour bar to the right hand side.}.
\end{center}
\end{figure}

We improve the algorithm in several ways: firstly instead of just 30
iterations, 
we allow a maximum of 100. This helps the program achieve convergence at some
parameter points. Secondly, we specify that the predicted value of
$M_Z^{pole}$ must be close enough to the empirical central value for
convergence to be possible. Thus, the convergence criteria are enlarged to
include the constraint that $|M_Z^{pole}/M_Z^{cent}-1|<\epsilon$ within the
first 10 iterations, or $|M_Z^{pole}/M_Z^{cent}-1|<10 \epsilon$ after the
first 10 iterations. Thirdly,
there are some sub iterations (for example for the
solution of Eq.~\ref{higgsMin} for $\mu(M_{SUSY})$ once loop corrections are
included) which are required to converge to much higher accuracy, given the 
sensitivity of each iteration to the underlying parameters.
Finally, we further smooth the iteration of $\mu(M_{SUSY})$ if the parameter
point is proving difficult. Between iteration 11 and 20, the average of 
the value of $\mu_{old}$ from the previous iteration and the one predicted by
the EWSB conditions acting on the current iteration $\mu_{ewsb}$ is
taken. Between iteration 21 and 30,  
{\tt SOFTSUSY3.3.6} takes $\mu(M_{SUSY})=0.2 \mu_{ewsb} + 0.8 \mu_{old}$,
whereas after iteration 31, it takes $\mu(M_{SUSY})=0.1 \mu_{ewsb} + 0.9
\mu_{old}$. We find that this helps dampen the oscillations apparent in the
points that exhibit poor convergence properties such as the 
`bad convergence' line in
Fig.~\ref{fig:points}. 
The resulting improvement in convergence is shown in
Fig.~\ref{fig:scanImp}. The region of no convergence has significantly
diminished, and 
become much more regular as compared to Fig.~\ref{fig:scan}a. Also, in the
region of successful EWSB, $M_Z$ is predicted to be much closer to
$M_Z^{cent}$, signalling a good solution to the EWSB conditions. 
These improvements are implemented in an improved version of {\tt
  SOFTSUSY} (version {\tt 3.3.6})~\cite{Allanach:2001kg}. 

\section{Summary and Conclusions \label{sec:sum}}

We have studied the uncertainties associated with the boundary of viable
parameter space in constrained high-scale SUSY mediation models of the MSSM
coming from successful EWSB\@. Even with the current 
accurate measurements of $m_t$, there is an enormous 2 TeV variation in the
CMSSM boundary's
position as measured by the soft SUSY breaking scalar mass input $m_0$. 
There is
also a theoretical uncertainty associated with higher order corrections to
the top mass, equivalent to around
0.8 GeV. Uncertainties on $m_t$ translate into an uncertainty on $h_t$, which
the RGEs are very sensitive to, and to which, ultimately, the EWSB region
becomes uncertain. There is therefore no concomitant EWSB uncertainty
in 
simplified models or other low-scale models such as the phenomenological MSSM\@.
As well as providing an uncertainty on the boundary
of viable parameter space, there is also, for a given squark and gluino mass
(assuming a particular high-scale model),
an uncertainty on the lightest neutralino mass which results in a theoretical
uncertainty upon cut efficiencies. Currently, these sources of uncertainty
are {\em not}\/ taken into account when interpreting experimental SUSY
searches in terms of particular high-scale models, and should be.
However, we 
have shown that they are significant and so must be taken into 
account for any robust interpretation of data in terms of some constrained
model of SUSY breaking at a high scale. The significance of the uncertainties
is likely to increase as the centre of mass energy of the collider increases,
since the probed region covers more of the uncertain region of parameter
space. 
In the CMSSM in particular, such uncertainties should
not have a significant impact on SUSY discovery, since the uncertainties occur
in a 
r\'{e}gime of large scalar masses, so discovery is based upon the production
of the other lighter sparticle states.  But they do have a large impact on
whether the CMSSM is a viable model or not (as opposed to some competing SUSY
breaking mediation model, for instance).

We now provide recommendations for those wishing to present
robust interpretations of searches
for supersymmetric particles, which one should apply 
when considering high-scale and constrained models of
SUSY breaking such as the CMSSM:
\begin{enumerate}
\item Generate default spectra using using a $m_t$ input value that is
  2$\sigma$ 
  plus a theoretical uncertainty (currently 0.8 GeV) 
{\em higher}\/ than the central value. This ensures that one is not erroneously
ruling out a region of parameter space for not breaking electroweak symmetry 
correctly.
\item Include a similar parametric variation of $m_t$ when estimating
  uncertainties on the interpretation of any search in the region close to the
  EWSB boundary, since it can change the kinematics of SUSY cascade chain
  decays, and therefore the efficiencies. 
  It is intended that the theoretical uncertainty will be
  reduced in the future by including higher order effects from
  Refs.~\cite{Ferreira:1996ug,Bednyakov:2002sf}   in {\tt SOFTSUSY}.
\item Use a recent version of {\tt SOFTSUSY}. 
In version {\tt 3.3.6}, we have improved the numerical convergence of {\tt
  SOFTSUSY} 
near the boundary of EWSB\@. We have also improved 
the diagnosis of an accurate solution of the RGEs and boundary conditions by
using the difference between the value of the $Z$ mass predicted and the
experimental value. 
\end{enumerate}
We hope  that ATLAS and CMS, as well as others, will adopt these
recommendations as well as the Les Houches
Recommendations~\cite{Kraml:2012sg}, leading to more 
robust interpretations of the searches.  

\section*{Acknowledgements}
This work has been partially supported by STFC\@. We thank other members of the
Cambridge SUSY working group for helpful comments.

\bibliographystyle{JHEP-2}
\bibliography{a}

\end{document}